\documentclass[aps,amsfonts,nofootinbib,onecolumn, superscriptaddress]{revtex4}
\usepackage{amsmath}
\usepackage{graphicx}
\usepackage{subfigure}
\usepackage{bm}

\newcommand{\beq}{\begin{equation}}
\newcommand{\eeq}{\end{equation}}
\newcommand{\bea}{\begin{eqnarray}}
\newcommand{\eea}{\end{eqnarray}}
\newcommand{\dk}{{\int \frac{\d^3 \vc{k}}{(2\pi)^{3/2}}}}

\newcommand{\tet}{{\tilde{\eta}}}

\newcommand{\vc}[1]{{\textbf{\em #1}}}

\newcommand{\equa}[1]{\begin{align} #1 \end{align}}
\newcommand{\pmtrx}[1]{\begin{pmatrix} #1 \end{pmatrix}}
\newcommand{\lh}{\left(}
\newcommand{\rh}{\right)}
\newcommand{\der}{\partial}
\renewcommand{\d}{\mathrm{d}}
\newcommand{\non}{\nonumber}
\DeclareMathSymbol{\mg}{\mathrel}{symbols}{"1D}
\newcommand{\ml}{\ll}

\newcommand{\half}{\frac{1}{2}}
\newcommand{\inv}{^{-1}}

\newcommand{\siml}{\raisebox{-.1ex}{\renewcommand{\arraystretch}{0.3}
$\begin{array}{@{}c} \scriptstyle < \\ \scriptstyle \sim
\end{array}$}}
\newcommand{\simg}{\raisebox{-.1ex}{\renewcommand{\arraystretch}{0.3}
$\begin{array}{@{}c} \scriptstyle > \\ \scriptstyle \sim
\end{array}$}}

\newcommand{\ga}{\alpha}
\newcommand{\gb}{\beta}
\renewcommand{\gg}{\gamma}
\newcommand{\gd}{\delta}

\newcommand{\gz}{\zeta}

\newcommand{\gth}{\theta}

\newcommand{\gk}{\kappa}

\newcommand{\gm}{\mu}
\newcommand{\gn}{\nu}

\newcommand{\gf}{\phi}

\newcommand{\gc}{\chi}
\newcommand{\gps}{\psi}
\newcommand{\go}{\omega}

\newcommand{\gG}{\Gamma}
\newcommand{\gD}{\Delta}
\newcommand{\gTh}{\Theta}

\newcommand{\gX}{\Xi}
\newcommand{\gP}{\varPi}

\newcommand{\gF}{\Phi}

\newcommand{\cD}{{\mathcal D}}

\newcommand{\cJ}{{\mathcal J}}

\newcommand{\cL}{{\mathcal L}}

\newcommand{\cO}{{\mathcal O}}

\newcommand{\cS}{{\mathcal S}}
\newcommand{\cT}{{\mathcal T}}

\newcommand{\cW}{{\mathcal{W}}}

\newcommand{\tb}{{\tilde b}}

\newcommand{\tf}{{\tilde f}}

\newcommand{\tB}{{\tilde B}}

\newcommand{\tge}{{\tilde\epsilon}}

\newcommand{\tgz}{{\tilde\zeta}}
\newcommand{\tget}{{\tilde\eta}}

\newcommand{\tgx}{{\tilde\xi}}

\newcommand{\bA}{{\bar A}}

\newcommand{\bC}{{\bar C}}

\newcommand{\bX}{{\bar X}}

\begin{document}

\title{Large non-Gaussianity in multiple-field inflation}

\author{G.I.~Rigopoulos}

\affiliation{Institute for Theoretical Physics, Utrecht University,\\
Postbus 80.195, 3508 TD Utrecht, The Netherlands}

\author{E.P.S.~Shellard}
\author{B.J.W.~van Tent}

\affiliation{Department of Applied Mathematics and Theoretical Physics,
Centre for Mathematical Sciences,\\
University of Cambridge,
Wilberforce Road, Cambridge CB3 0WA, United Kingdom}

\begin{abstract}

\noindent We investigate non-Gaussianity in general multiple-field
inflation using the formalism we developed in earlier papers. We
use a perturbative expansion of the non-linear equations to
calculate the three-point correlator of the curvature perturbation
analytically. We derive a general expression that involves only a
time integral over background and linear perturbation quantities.
We work out this expression explicitly for the two-field slow-roll
case, and find that non-Gaussianity can be orders of magnitude
larger than in the single-field case. In particular, the
bispectrum divided by the square of the power spectrum can easily
be of $\cO(1$--$10)$, depending on the model. Our result also
shows the explicit momentum dependence of the bispectrum. This
conclusion of large non-Gaussianity is confirmed in a semi-analytic 
slow-roll investigation of a simple quadratic two-field model.

\end{abstract}

\maketitle

\section{Introduction}

The assumption that inflationary fluctuations are Gaussian is a
good starting point for the study of cosmological perturbations,
but it is only true to linear order in perturbation theory. Since
gravity is inherently non-linear, and most inflation models have
\mbox{(self-)}interacting potentials, non-linearity must be
present at some level in all inflation models. Hence the issue is
not whether inflation is non-Gaussian, but how large the
non-Gaussianity is. With increasingly precise CMB data becoming
available in the near future from the WMAP and Planck satellites
and other experiments, we might well hope to detect this
non-Gaussianity. This would offer us another key observable to
help constrain or confirm specific inflation models and the
underlying high-energy theories from which they are derived. As a
rough order of magnitude estimate, we note that non-Gaussianity
will be detectable by Planck if the bispectrum (the Fourier
transform of the three-point correlator) is of the order of the
square of the power spectrum \cite{komatsu}.

It follows that to compute the predicted amount of non-Gaussianity
in specific inflation models we need to go beyond linear-order
perturbation theory. In \cite{gp2,formalism} we introduced a new
formalism to deal with the non-linearity during inflation. We will
not again summarise the other work dealing with this subject,
references for which can be found in \cite{formalism} or a recent
review \cite{ngreview}. Our formalism is distinguished by being
based on a system of fully non-linear equations for long
wavelengths, while stochastic sources take into account the
continuous influx of short-wavelength fluctuations into the 
long-wavelength system as the inflationary comoving horizon shrinks.
The variables used incorporate both scalar metric and matter
perturbations self-consistently and they are invariant under
changes of time slicing in the long-wavelength limit.

The advantages of our method are threefold: {\em (i)} it is
physically intuitive and relatively simple to use for quantitative
analytic and semi-analytic calculations; {\em (ii)} it is valid in
a very general multiple-field inflation setting, which includes
the possibility of a non-trivial field metric; and {\em (iii)} it
is well-suited for direct numerical implementation. The first
point was already demonstrated in \cite{sf}, where we computed the
non-Gaussianity in single-field slow-roll inflation, while the
third point is the subject of a forthcoming paper \cite{RSvTnum}.
The present paper is dedicated to the exploration of the second
point, as well as the first. 

In \cite{sf} we found, confirming
what was known in the literature beforehand (see e.g.\
\cite{maldacena}), that non-Gaussianity in the single-field case
is too small to be realistically observable, because it is
suppressed by slow-roll factors (actually the scalar spectral
index $n-1$). However, there have been long-standing claims in the
literature (see e.g.\ \cite{salopek, uzan}) that specific
multiple-field models can, in principle, create significant
non-Gaussianity.  Indeed, there has been growing recent interest
in models which can produce large primordial non-Gaussianity
\cite{large ng}. A feature shared by most of these models though
is that this non-Gaussianity involves some mechanism operating
\emph{after} inflation, usually (p)reheating or later domination
of a curvaton field. In this paper we investigate for the first
time general multiple-field inflation, not just specific models,
presenting a method by which to accurately calculate the resulting
three-point correlator \emph{during} inflation. We find that it is
possible to get significant primordial non-Gaussianity without
invoking some post-inflationary mechanism even for the simplest
two-field quadratic potential. 

The key mechanism for the
production of this large non-Gaussianity is the superhorizon
influence of isocurvature perturbations on the adiabatic mode. The
former feed into the latter when the background follows a curved
trajectory in field space. Note that the example studied in
section~\ref{realmodel} illustrates that there is no need for the
potential to be interacting. Our aim is to push forward towards a
tractable non-Gaussian methodology for the new era of precision
cosmology which confronts us.

This work is organised as follows. In section~\ref{setupsec} we
give the equations from \cite{formalism} that are used as the
starting point for the present investigations. In
section~\ref{generalsec} we then derive the general solution for
the relevant quantities in multiple-field inflation, culminating
in a general expression for the three-point correlator of the
adiabatic component of the curvature perturbation, without any
slow-roll approximations. This integral solution  - equation
(\ref{generalfkkreal}) - is a very useful calculational tool
because it gives the three-point correlator entirely in terms of
background quantities and linear perturbation quantities at
horizon crossing. In section~\ref{SRapproxsec} we make a
leading-order slow-roll approximation to work out the various
contributions in the general solution more explicitly. Finally in
section~\ref{expl2fsec} we calculate the bispectrum in an analytic
treatment of the two-field case with constant slow-roll
parameters. We find that the result can be orders of magnitude
larger than for single-field inflation. This result is confirmed
with a semi-analytic slow-roll calculation of an explicit model with a
quadratic potential in section \ref{realmodel}. Our method yields
the full momentum dependence, not just an overall magnitude, and
we find that there can be a difference of the order of a few
between opposite extreme momentum limits. We conclude in
section~\ref{conclsec}. Parts of this paper, in particular 
section~\ref{expl2fsec}, are rather technical, so some readers might 
be interested in referring to \cite{mf2} first, which contains a 
simplified derivation of only the dominant non-Gaussian contributions in
multiple-field inflation, along with a summarised discussion.

\section{Multiple-field setup}
\label{setupsec}

Since in this paper we are explicitly working out the general
non-linear multiple-field formalism of \cite{formalism}, we refer
the reader to that paper for derivations and more details of the
initial equations. Here we just briefly describe the context and
give the relevant equations and definitions to be used as starting
point for further calculations.

We start from a completely general inflation model, with an
arbitrary number of scalar fields $\gf^A$ (where $A$ labels the
different fields) and a potential $V(\gf^A)$ with arbitrary
interactions. We also allow for the possibility of a non-trivial
field manifold with field metric $G_{AB}$. We will consider only
scalar modes and make the long-wavelength approximation (i.e.\
consider only wavelengths larger than the Hubble radius $1/(aH)$,
where second-order spatial gradients can be neglected compared to
time derivatives)\footnote{Formally this corresponds with taking
only the leading-order terms in the gradient expansion. We expect
higher-order terms to be subdominant on long wavelengths during
inflation, but this statement has only been rigorously verified at
the linear level. A calculation to higher order in spatial
gradients, or, even better, a full proof of convergence of the
expansion, would be desirable. See \cite{gradient} for more
details on the validity of the gradient expansion beyond linear
theory.}. The spacetime metric $g_{\gm\gn}$ and matter Lagrangean
are given by
    \beq
    \d s^2=-N^2(t,\vc{x})\d t^2+a^2(t,\vc{x})\d\vc{x}^2,
    \qquad\qquad
    \cL_\mathrm{m} = - \half g^{\mu\nu} \der_\gm \gf^A G_{AB} \der_\gn \gf^B
    - V(\gf^A),
    \eeq
with $a$ the local scale factor and $N$ the lapse function. The
local expansion rate is defined as $H \equiv \dot{a}/(N a)$, where
the dot denotes a derivative with respect to $t$. The proper field
velocity is $\gP^A \equiv \dot{\gf}^A/N$, with length $\gP$. We
also define local slow-roll parameters as
    \beq
    \tge(t,\vc{x}) \equiv \frac{\gk^2 \gP^2}{2 H^2} \,,
    \qquad\qquad
    \tget^A(t,\vc{x}) \equiv - \frac{3 H \gP^A
    + G^{AB} \der_B V}{H\gP}\,,
    \qquad\qquad
    \tgx^A(t,\vc{x}) \equiv - \frac{\cD_B \der^A V}{H^2} \frac{\gP^B}{\gP}
    + 3 \tge \, \frac{\gP^A}{\gP} - 3 \tget^A,
    \label{slowrollpar}
    \eeq
where $\gk^2 \equiv 8\pi G = 8\pi/m^2_\mathrm{pl}$ and $\cD_B$ is
a covariant derivative with respect to the field $\gf^B$. For the
first part of the paper we will not make a slow-roll
approximation, and consider these definitions as just a short-hand
notation. When we do make this approximation, from
section~\ref{SRapproxsec}, $\tge$ and $\tget^A$ are first order in
slow roll, while $\tgx^A$ is second order. Finally, we choose the
gauge where
    \beq\label{time}
    t=\ln(aH) \qquad\Leftrightarrow\qquad
    NH=(1-\tge)^{-1}.
    \eeq
In this gauge horizon exit of a mode, $k=aH$, occurs
simultaneously for all spatial points and calculations are
simpler.

We will make use of a preferred basis in field space, defined as
follows. The first basis vector $e_1^A$ is the direction of the
field velocity. Next, $e_2^A$ is defined as the direction of that
part of the field acceleration that is perpendicular to $e_1^A$.
One continues this orthogonalisation process with higher
derivatives until a complete basis is found. Explicit expressions
can be found in \cite{formalism}, here we only give $e_1^A \equiv
\gP^A/\gP$. Now one can take components of vectors in this basis
and we define, for example for $\gz_i^A$ (defined below in
(\ref{gi_var})) and $\tget^A$:
    \beq\label{Lmdef}
    \gz_i^m \equiv e_{m\,A} \gz_i^A,
    \qquad\qquad
    \tget^\parallel \equiv e_1^A \tget_A,
    \qquad\qquad
    \tget^\perp \equiv e_2^A \tget_A.
    \eeq
Note that, unlike for the index $A$, there is no difference
between upper and lower indices for the $m$. By construction there
are no other components of $\tget^A$, so that one can write
$\tget^\perp = |\tget^A - \tget^\parallel e_1^A|$. We also define
    \beq
    Z_{mn} \equiv \frac{1}{NH} \, e_{m\,A} \cD_t e_n^A,
    \eeq
where $\cD_t$ is the covariant time derivative containing the
connection of the field manifold. $Z_{mn}$ is antisymmetric and
only non-zero just above and below the diagonal, and first order
in slow roll. Its explicit form in terms of slow-roll parameters
can be found in \cite{vantent}; here we only need that $Z_{12} = -
Z_{21} = -\tget^\perp$.

As discussed in \cite{gp2,formalism} it is useful to work with the
following combination of spatial gradients to describe the fully
non-linear inhomogeneities:
    \beq\label{gi_var}
    \gz_i^A(t,\vc{x}) \equiv e_1^A(t,\vc{x}) \der_i \ln a(t,\vc{x})
    - \frac{\gk}{\sqrt{2\tge(t,\vc{x})}} \, \der_i \gf^A(t,\vc{x})
    \qquad\Leftrightarrow\qquad
    \gz_i^m \equiv \gd_{m1} \der_i \ln a
    - \frac{\gk}{\sqrt{2\tge}} \,
    e_{m\,A} \der_i \gf^A,
    \eeq
which is invariant under changes of time slicing, up to
second-order spatial gradients \cite{gp, formalism}. Note that,
when linearised, $\gz_i^1$ (the $m$=1 component of $\gz_i^m$) is
the spatial gradient of the well-known $\gz$ from the literature,
the curvature perturbation. In \cite{formalism} we derived a fully
non-linear equation of motion for $\gz_i^m$ without any slow-roll
approximations:
    \beq\label{basic}
    \left\{ \begin{array}{l}
    \displaystyle
    \dot{\gz}_i^m - \gth_i^m = \cS_i^m\\
    \displaystyle
    \dot{\gth}_i^m + \lh \frac{(3 - 2\tge + 2\tget^\parallel - 3\tge^2
    - 4\tge\tget^\parallel)\gd_{mn}}{(1-\tge)^2} + \frac{2Z_{mn}}{1-\tge} \rh
    \gth_i^n + \frac{\gX_{mn}}{(1-\tge)^2} \, \gz_i^n = \cJ_i^m
    \end{array} \right.
    \eeq
where $\gth_i^m$ is the velocity corresponding with $\gz_i^m$ and
    \bea
    \gX_{mn} & \equiv & \frac{V_{mn}}{H^2} - \frac{2\tge}{\gk^2} R_{m11n}
    + (1-\tge)\dot{Z}_{mn} + Z_{mp}Z_{pn}
    + \lh 3-2\tge+2\tget^\parallel-\tge^2-2\tge\tget^\parallel \rh
    \frac{Z_{mn}}{1-\tge} \\
    && + \lh 3\tge+3\tget^\parallel+2\tge^2+4\tge\tget^\parallel+(\tget^\perp)^2
    +\tgx^\parallel \rh \gd_{mn}
    - 2\tge \lh (3+\tge+2\tget^\parallel)\gd_{m1}\gd_{n1} 
    + \tget^\perp (\gd_{m1}\gd_{n2}+\gd_{m2}\gd_{n1}) \rh, \non
    \eea
where $V_{mn} \equiv e_m^A (\cD_B \der_A V) e_n^B$ and $R_{m11n}
\equiv e_m^A R_{ABCD} e_1^B e_1^C e_n^D$ with $R^A_{\;BCD}$ the
curvature tensor of the field manifold. Although for the first
part of the paper we will not make a slow-roll approximation, we
give here immediately the leading-order slow-roll approximation of
(\ref{basic}), which we will be using in the second part, to show
that things simplify considerably in that case:
    \beq\label{basicSR}
    \left\{ \begin{array}{l}
    \displaystyle
    \dot{\gz}_i^m - \gth_i^m = \cS_i^m\\
    \displaystyle
    \dot{\gth}_i^m + 3 \gth_i^m + \biggl( \frac{V_{mn}}{H^2} + 3 Z_{mn}
    + 3(\tge+\tget^\parallel)\gd_{mn}
    - 6\tge\,\gd_{m1}\gd_{n1} \biggr) \gz_i^n = \cJ_i^m
    \end{array} \right.
    \eeq
The stochastic source terms $\cS_i^m$ and $\cJ_i^m$ are given by
    \bea
    \cS_i^m & = & -\frac{\gk}{a\sqrt{2\tge}} \dk \, \dot{\cW}(k)
    Q_{mn}^\mathrm{lin}(k) \ga_n(\vc{k}) \,
    \mathrm{i} k_i \mathrm{e}^{\mathrm{i} \vc{k}\cdot\vc{x}}
    + \mathrm{c.c.}, \non\\
    \cJ_i^m & = & -\frac{\gk}{a\sqrt{2\tge}} \dk \, \dot{\cW}(k)
    \left[ \dot{Q}_{mn}^\mathrm{lin}(k) \gd_{np}
    - Q_{mn}^\mathrm{lin}(k) \, 
    \frac{(1+\tge+\tget^\parallel)\gd_{np} + Z_{np}}{1-\tge} \right] 
    \ga_p(\vc{k}) \,
    \mathrm{i} k_i \mathrm{e}^{\mathrm{i} \vc{k}\cdot\vc{x}}
    + \mathrm{c.c.},
    \label{sources}
    \eea
where c.c.\ denotes the complex conjugate. The perturbation
quantity $Q_{mn}^\mathrm{lin}$ is the solution from linear theory
for the multiple-field generalisation of the Sasaki-Mukhanov
variable $Q \equiv - a \sqrt{2 \tge} \, \gz / \gk$. It can be
computed exactly numerically, or analytically within the slow-roll
approximation \cite{vantent}. The $\ga_m(\vc{k})$ are a set of
Gaussian complex random numbers satisfying
    \beq\label{Gcrn}
    \langle \ga_m^{}(\vc{k}) \ga_n^*(\vc{k}')\rangle
    =\gd_{mn}\gd^3\!(\vc{k}-\vc{k}'),
    \qquad\qquad
    \langle \ga_m(\vc{k}) \ga_n(\vc{k}')\rangle=0.
    \eeq
The quantity $\cW(k)$ is the Fourier transform of an appropriate
smoothing window function which cuts off modes with wavelengths
smaller than the Hubble radius; we choose it to be a Gaussian with
smoothing length $R\equiv c/(aH) = c \, \mathrm{e}^{-t}$, where
\mbox{$c\approx\,$3--5:}
    \beq
    \cW(k) = \mathrm{e}^{-k^2 R^2 /2}
    \qquad\Rightarrow\qquad
    \dot{\cW}(k) = k^2 R^2 \mathrm{e}^{-k^2 R^2 /2}.
    \eeq
Since $\gz_i^m$ and $\gth_i^m$ are smoothed long-wavelength
variables, the appropriate initial conditions are that they should
be zero at early times when all the modes are sub-horizon. Hence,
    \beq
    \lim_{t \rightarrow -\infty} \gz_i^m = 0,
    \qquad\qquad
    \lim_{t \rightarrow -\infty} \gth_i^m = 0.
    \eeq

A key aspect of the system (\ref{basic}) or (\ref{basicSR}) is
that it is fully non-linear. All functions in the coefficients on
the left-hand side of the equation, like $\tge(t,\vc{x})$, and in
the sources on the right-hand side are inhomogeneous and depend on
$\gz_i^m$ and $\gth_i^m$ via a basic set of three constraint
equations:
    \bea
    \der_i \ln a & = & - \der_i \ln H
    = - \frac{\tge}{1-\tge} \, e_{1\,A} \gz_i^A,
    \label{constr1}\\
    \der_i \gf^A & = & - \frac{\sqrt{2\tge}}{\gk} \lh \gd^A_B
    + \frac{\tge}{1-\tge} \, e_1^A e_{1\,B} \rh \gz_i^B,
    \label{constr2}\\
    \cD_i \gP^A & = & - \frac{\sqrt{2\tge}}{\gk} \, H \biggl[ (1-\tge) \gth_i^A
    + \lh (\tge + \tget^\parallel)\gd^A_B - \tge \, e_1^A e_{1\,B}
    + \frac{\tge}{1-\tge} \, \tget^A e_{1\,B} \rh \gz_i^B \biggr].
    \label{constr3}
    \eea
Using only these three constraints one can compute the spatial
derivative of all relevant quantities, keeping in mind that
$\gth_i^m = e^{\;}_{m\,A} \gth_i^A - (1-\tge)\inv Z_{mn} \gz_i^n$.
Note that in our gauge $\cW$ depends on $t$ only and does not get
any non-linear contributions.

\section{General analytic solution}
\label{generalsec}

In this section we investigate how to solve the system
(\ref{basic}) analytically and give formal expressions for the
solution. In the next sections we will investigate cases where we
can determine the solution more explicitly. We start by rewriting
the system (\ref{basic}) into a single vector equation:
    \beq\label{basic_sym}
    \dot{v}_{i\,a}(t,\vc{x}) + A_{ab}(t,\vc{x}) v_{i\,b}(t,\vc{x})
    = b_{i\,a}(t,\vc{x}),
    \qquad
    \lim_{t\rightarrow-\infty} v_{i\,a} = 0,
    \qquad
    v_i \equiv \pmtrx{\gz_i^1\\\gth_i^1\\\gz_i^2\\\gth_i^2\\\vdots},
    \qquad
    b_i \equiv \pmtrx{\cS_i^1\\\cJ_i^1\\\cS_i^2\\\cJ_i^2\\\vdots}\,.
    \eeq
Here the indices $a,b,\ldots$ label the components within this
$2N$-dimensional space (with $N$ the number of fields). The matrix
$A$ can be read off from (\ref{basic}) and has the following form:
$A_{2m-1,2m}=-1$, $A_{2m,2n}=\gTh_{mn}$ and
$A_{2m,2n-1}=\gX_{mn}/(1-\tge)^2$, where $\gTh_{mn}$ is the matrix
between parentheses in the second equation of (\ref{basic}) and
$m,n=1,2,\ldots,N$. All other entries of $A$ are zero.

Equation (\ref{basic_sym}) is non-linear since the matrix
$A(t,\vc{x})$ and the the source term $b_i(t,\vc{x})$ are
inhomogeneous functions in space and depend on the $v_i$ through
(\ref{constr1})--(\ref{constr3}). It can be solved perturbatively
as an infinite hierarchy of linear equations with known source
terms at each order (see also \cite{formalism}). We expand the
relevant quantities as
    \beq
    v_i=v^{(1)}_i+v^{(2)}_i+\ldots\,,
    \qquad
    b_i=b_i^{(1)}+b_i^{(2)}+\ldots\,,
    \qquad
    A(t,\vc{x}) = A^{(0)}(t) + A^{(1)}(t,\vc{x}) + A^{(2)}(t,\vc{x}) + \ldots
    \eeq
Then the equation of motion for $v_i^{(m)}$ is
    \beq\label{vimeq}
    \dot{v}^{(m)}_{i\,a}(t,\vc{x}) + A^{(0)}_{ab}(t)
    v^{(m)}_{i\,b}(t,\vc{x}) = \tb^{(m)}_{i\,a}(t,\vc{x}), \qquad
    \lim_{t\rightarrow-\infty} v_{i\,a}^{(m)} = 0, \qquad
    \tb^{(m)}_{i\,a} \equiv b^{(m)}_{i\,a} -
    \sum^{m-1}_{j=1}A^{(m-j)}_{ab}v^{(j)}_{i\,b}.
    \eeq
Let us recapitulate the meaning of the various indices, to avoid
confusion. The index $i=1,2,3$ labels the components of spatial
vectors. The indices $A,B,\ldots=1,\ldots,N$ label components in
field space. These indices will not occur in the rest of the
paper, since they have been replaced by the indices
$m,n,\ldots=1,\ldots,N$ that label components in field space
within the special basis as defined in (\ref{Lmdef}). Next, the
indices $a,b,\ldots=1,\ldots,2N$ label components within the
$2N$-dimensional space consisting of both $\gz$ and $\gth$ as
defined in (\ref{basic_sym}). Finally there are the labels within
parentheses that denote the order in the perturbative expansion
defined above. Only with the $i$ and $A,B,\ldots$ is there a
difference between upper and lower indices.

We now show that the source term $\tb^{(m)}_i$ is known from the
solutions for $v_i$ up to order $(m-1)$. The equation for
$v_i^{(1)}$ is linear by construction: $A^{(0)}$ depends only on
the homogeneous background quantities, and the only $\vc{x}$
dependence in $b_i^{(1)}$ is in the
$\mathrm{e}^{\mathrm{i}\vc{k}\cdot\vc{x}}$, for the rest it
depends on homogeneous background quantities. All of these are in
the end functions of just $H$, $\gf^A$, and $\gP^A$ via their
definitions. To go beyond linear order all these background
quantities are perturbed as follows ($C$ stands for any of the
quantities to be perturbed, for example $\tge$, $\tget^\perp$,
etc.):
    \beq\label{expansion}
    C(t,\vc{x}) \:=\: C^{(0)}(t) + C^{(1)}(t,\vc{x}) + \ldots
    \:=\: C^{(0)} + \der^{-2} \der^i (\der_i C)^{(1)} + \ldots
    \:=\: C^{(0)} + \bC_a^{(0)} \der^{-2} \der^i v_{i\,a}^{(1)} + \ldots
    \eeq
where we use (\ref{constr1})--(\ref{constr3}) to compute $\der_i
C$ and $\bC^{(0)}$ is some homogeneous (space-independent) vector
that is the result of that calculation. Next, to compute $C^{(2)}$
one simply repeats this process with the vector $\bC$, and
continues in this way order by order (of course there is also a
$\bC^{(0)}_a \der^{-2} \der^i v_{i\,a}^{(2)}$ term at second
order, etc.). Then it is easy to see that $\tb_i^{(m)}$ depends
only on $v_i^{(1)},\ldots,v_i^{(m-1)}$, and hence is a known
quantity at each order.

The solution of equation (\ref{vimeq}) for $v_i^{(m)}$ can be
written as
    \beq\label{solviG}
    v_{i\,a}^{(m)}(t,\vc{x})=\int_{-\infty}^t \d t' \, G_{ab}(t,t')
    \,\tb_{i\,b}^{(m)}(t',\vc{x}),
    \eeq
with the Green's function $G_{ab}(t,t')$ satisfying\footnote{To be
precise, the Green's function is actually defined as the solution
of (\ref{Green}) with $\gd(t-t')$ on the right-hand side instead
of zero. The solution is then a step function times what we call
the Green's function. This step function has been taken into
account by changing the upper limit of the integral in
(\ref{solviG}) from $\infty$ to $t$.}
    \beq\label{Green}
    \frac{\d}{\d t}\, G_{ab}(t,t') + A^{(0)}_{ac}(t) G_{cb}(t,t') = 0,
    \qquad\qquad
    \lim_{t\rightarrow t'} G_{ab}(t,t') = \gd_{ab}.
    \eeq
It is important to note that this Green's function is homogeneous,
a solution of the background equation. It has to be computed only
once, and can then be used to calculate the solution at each order
using the different source terms as in (\ref{solviG}). We write
explicitly for the first two orders:
    \bea
    b_{i\,a}^{(1)}(t,\vc{x}) & = & \dk \, \dot{\cW}(k,t)
    X_{am}^{(1)}(k,t) \ga_m(\vc{k}) \,
    \mathrm{i}k_i \,\mathrm{e}^{\mathrm{i} \vc{k}\cdot\vc{x}}
    + \mathrm{c.c.},
    \non\\
    b_{i\,a}^{(2)}(t,\vc{x}) & = & \lh\der^{-2}
    \der^i v_{i\,c}^{(1)}(t,\vc{x})\rh
    \dk \, \dot{\cW}(k,t) \bX_{amc}^{(1)}(k,t) \ga_m(\vc{k}) \,
    \mathrm{i}k_i \,\mathrm{e}^{\mathrm{i} \vc{k}\cdot\vc{x}}
    + \mathrm{c.c.}
    \label{bia2}
    \eea
Comparison with (\ref{sources}) shows that $X_{am}$ is given by
the following equations:
    \beq\label{tXdef}
    X_{2n-1,m} = -\frac{\gk}{a\sqrt{2\tge}} \, Q_{nm}^\mathrm{lin},
    \qquad\qquad
    X_{2n,m} = -\frac{\gk}{a\sqrt{2\tge}}
    \left[ \dot{Q}_{nm}^\mathrm{lin} - Q_{np}^\mathrm{lin} \,
    \frac{(1+\tge+\tget^\parallel)\gd_{pn} + Z_{pn}}{1-\tge}\,
    \right] .
    \eeq
The quantity $\bX_{amc}$ is derived from $X_{am}$ as in
(\ref{expansion}), but in addition it also contains the perturbation of
the basis vector $e_m$ inside the $\ga_m$. In the same way we define
$A^{(1)}_{ab}(t,\vc{x}) = \bA^{(0)}_{abc}(t) \der^{-2} \der^i
v^{(1)}_{i\,c}(t,\vc{x})$.

Using the solution (\ref{solviG}), valid at each order, and the
relations (\ref{Gcrn}) to compute the averages, it is now
straightforward to write down the general expressions for the
two-point and three-point correlators of the adiabatic ($m=1$)
component of $\gz^m \equiv \der^{-2} \der^i \gz_i^m$, which is the
$a=1$ component of $v_{i\,a}$, or rather their Fourier transforms,
the power spectrum and the bispectrum. Making use of the
short-hand notation
    \beq\label{zetaamp}
    v^{(1)}_{am}(k,t) \equiv \int_{-\infty}^t \d t' \,
    G_{ab}(t,t') \dot{\cW}(k,t') X_{bm}^{(1)}(k,t'),
    \eeq
we find for the power spectrum:
    \beq\label{powerspec}
    \left\langle |\gz^{(1)\,1}(k,t)|^2 \right\rangle
    = v^{(1)}_{1m}(k,t) v^{(1)\,*}_{1m}(k,t) + \mathrm{c.c.}
    \eeq
We emphasise here that the quantity $v^{(1)}_{am}$ defined by
(\ref{zetaamp}) is simply made up of the linear
$\gz_{mn}^\mathrm{lin}$, $\gth_{mn}^\mathrm{lin}$ mode functions.
One needs to evaluate the Green's function $G_{ab}(t,t')$ and to
perform the integral (\ref{zetaamp}) when the linear source terms
$Q_{mn}^\mathrm{lin}$, $\dot{Q}_{mn}^\mathrm{lin}$ in
$X_{bm}^{(1)}$ are known only up to horizon crossing, $k \approx
aH$.  However, where analytic solutions are available for $k \ml
aH$, or in a fully numerical approach, we can dispense with the
integral (\ref{zetaamp}) by using the linear solution on
super-horizon scales.

To find the bispectrum the calculation is slightly longer. One
first has to compute $\gz^{(2)\,1}(t,\vc{x})$. As we noted in the
single-field case in \cite{sf}, $\langle \gz^{(2)\,1} \rangle$ is
indeterminate. To remove this ambiguity and also require that
perturbations have a zero average, we define $\tgz^m \equiv \gz^m
- \langle \gz^m \rangle$. Expanding
$\tgz^m=\tgz^{(1)\,m}+\tgz^{(2)\,m}$, the three-point correlator
will be a combination of the different permutations of $\langle
\tgz^{(2)\,1}(t,\vc{x}_1) \tgz^{(1)\,1}(t,\vc{x}_2)
\tgz^{(1)\,1}(t,\vc{x}_3) \rangle$, and the bispectrum is its
Fourier transform. The intermediate steps are given in more detail
in the explicit calculation in section~\ref{expl2fsec}; here we go
directly to the end result for the bispectrum of the adiabatic
component:
    \beq\label{generalbispec}
    \left\langle \tgz^1(t,\vc{x}_1) \tgz^1(t,\vc{x}_2) \tgz^1(t,\vc{x}_3)
    \right\rangle^{(2)}(\vc{k}_1,\vc{k}_2,\vc{k}_3)
    = (2\pi)^3 \gd^3(\vc{k}_1 + \vc{k}_2 + \vc{k}_3)
    \Bigl[ f(\vc{k}_1,\vc{k}_2) + f(\vc{k}_1,\vc{k}_3)
    + f(\vc{k}_2,\vc{k}_3) \Bigr]
    \eeq
with
    \equa{
    f(\vc{k},\vc{k}') \equiv \frac{k^2 + \vc{k}\cdot\vc{k}'}{|\vc{k}+\vc{k}'|^2}
    \, v^{(1)\,*}_{1m}(k,t) \int_{-\infty}^t \d t' \, & G_{1a}(t,t')
    \left[ \bX^{(1)}_{amc}(k,t') \dot{\cW}(k,t') - \bA^{(0)}_{abc}(t')
    v^{(1)}_{bm}(k,t') \right]
    \non\\
    & \times \left[ v^{(1)\,*}_{1n}(k',t) v^{(1)}_{cn}(k',t')
    + v^{(1)}_{1n}(k',t) v^{(1)\,*}_{cn}(k',t') \right]
    + \mathrm{c.c.} + \vc{k} \leftrightarrow \vc{k}'.
    \label{generalfkk}
    }
If $Q^\mathrm{lin}_{mn}$ is real, this simplifies to
    \equa{
    f(\vc{k},\vc{k}') \equiv \: & 4 \,
    \frac{k^2 + \vc{k}\cdot\vc{k}'}{|\vc{k}+\vc{k}'|^2}
    \, v^{(1)}_{1m}(k,t) v^{(1)}_{1n}(k',t)
    \non\\
    & \times \int_{-\infty}^t \d t' \, G_{1a}(t,t')
    \left[ \bX^{(1)}_{amc}(k,t') \dot{\cW}(k,t') - \bA^{(0)}_{abc}(t')
    v^{(1)}_{bm}(k,t') \right] v^{(1)}_{cn}(k',t')
    + \vc{k} \leftrightarrow \vc{k}'.
    \label{generalfkkreal}
    }
This integral expression is a key result of this paper. Using our
methodology, the three-point correlator with full momentum
dependence has been expressed as a single time integral over
quantities determined by the background model and the linear
perturbations, that is, respectively the matrix $\bA^{(0)}_{abc}$
and the solution $Q_{mn}^\mathrm{lin}$ embedded in
$\bX_{amc}^{(1)}$ (\ref{tXdef}) and $v_{am}^{(1)}$ (\ref{zetaamp})
(in both of which the background is also implicit). Of course, one
also has to find the Green's function $G_{ab}$ from (\ref{Green}),
but, like the equation for $Q^A_{\mathrm{lin}\,B}$ in
\cite{formalism}, this is a linear ordinary differential equation
for which there is no serious impediment to finding a numerical
solution, in cases where an analytic or semi-analytic solution is
unknown. The integral solution (\ref{generalfkkreal}), then,
demonstrates that the calculation of the three-point correlator is
straightforward and tractable. It is, in principle, similar to
calculations of the power spectrum, where accurate estimates can
be found from background quantities, for example, in the slow-roll
approximation. Here, we only have to supplement this with the
amplitudes of the linear perturbation mode functions
$Q_{mn}^\mathrm{lin}(k,t)$ and the closely related Green's
function $G_{ab}(t,t')$. In section~\ref{realmodel}, using this
methodology, we provide some quantitative semi-analytic results
for the bispectrum of a two-field inflation model with a quadratic
potential.

Before closing this section a final comment is in order. A
feature of (\ref{generalfkkreal}) which may at first sight cast
doubt on its utility for quantitative calculations is its apparent
dependence on the ad hoc choice for the functional form of the
window function $\cW(k)$. Closer inspection reveals that the
second term of (\ref{generalfkkreal}) (the $\bar{A}$ term) does
not depend on $\cW(k)$ for scales sufficiently larger than the
horizon. This is evident from the fact that (\ref{zetaamp}) is
simply the solution to linear theory smoothed on scales larger
than the horizon. Any properly normalised window function with
$\cW(k)\rightarrow 1$ for scales sufficiently larger than the
horizon will produce the same final answer. The $\bar{A}$ term
represents the non-linear evolution on superhorizon scales and, as
we show below, it describes an integrated effect which can lead to
large non-Gaussianity. In contrast, the $\bar{X}$ term arises
from perturbations around horizon crossing and may depend on
$\cW$. We find below that this term is localized around horizon
crossing and that it does not give rise to observationally interesting
effects. Section~\ref{Discuss} provides more discussion on
these points. In \cite{mf2} we explicitly show that taking a step
function instead of a Gaussian as window function does not change
the leading-order integrated effects.

\section{Slow-roll approximation}
\label{SRapproxsec}

The perturbation quantity $Q_{mn}^\mathrm{lin}$ can be computed
exactly numerically, or analytically within the slow-roll
approximation where all slow-roll parameters are assumed to
be smaller than unity. The latter was done in \cite{vantent} to
next-to-leading order in slow roll:\footnote{Compared with the
solution in \cite{vantent} there is an extra factor of
$1/\sqrt{2}$. It has to be introduced to take into account the
difference between the classical Gaussian random numbers $\ga$,
which have $\langle \ga \ga^* \rangle = \langle \ga^* \ga
\rangle$, and the quantum creation/annihilation operators
$\hat{a}^\dagger$, $\hat{a}$, which have $\langle \hat{a}^\dagger
\hat{a} \rangle = 0$. In \cite{sf} we introduced this factor of
$1/\sqrt{2}$ in the analogue of equation (\ref{sources}), which
leads of course to identical results.}
    \beq\label{linsol}
    Q_{mn}^\mathrm{lin} =  \frac{1}{2\sqrt{k}}
    \left[ E \lh \frac{c}{kR} \rh^{1+D} \right]_{mn},
    \eeq
where the matrices $D$ and $E$ are defined by
    \beq
    D_{mn} \equiv \tge \gd_{mn} + 2\tge \gd_{m1} \gd_{n1} - \frac{V_{mn}}{3H^2},
    \qquad\qquad
    E_{mn} \equiv (1-\tge)\gd_{mn} + \lh 2- \gg_E - \ln 2 \rh D_{mn},
    \eeq
with $\gg_E$ Euler's constant. Overall unitary factors that are
physically irrelevant have been omitted. Using this expression the
source terms are given by
    \bea
    \cS^m_i &=& -\frac{\gk}{a\sqrt{2\tge}} \dk \, \dot{\cW}(k)
    \, Q_{mn}^\mathrm{lin}(k) \ga_n(\vc{k}) \,
    \mathrm{i}k_i \,\mathrm{e}^{\mathrm{i} \vc{k}\cdot\vc{x}}
    +\mathrm{c.c.},
    \non\\
    \cJ^m_i &=& -\frac{\gk}{a\sqrt{2\tge}}
    \left[ D_{mn} - (2\tge+\tget^\parallel)\gd_{mn} - Z_{mn} \right] 
    \dk \, \dot{\cW}(k) \, Q_{np}^\mathrm{lin}(k) \ga_p(\vc{k}) \,
    \mathrm{i}k_i \, \mathrm{e}^{\mathrm{i} \vc{k}\cdot\vc{x}}
    +\mathrm{c.c.}
    \label{sourcesSR}
    \eea
Now when computing $\bX_{amc}^{(1)}$ as defined in (\ref{bia2}),
or any higher-order terms in the perturbative expansion, we would
in principle have to make the background quantities in
$Q_{mn}^\mathrm{lin}$ dependent on $\vc{x}$ and perturb them
according to (\ref{expansion}). However, from (\ref{linsol}) we
see that $Q_{mn}^\mathrm{lin}$ depends on $\vc{x}$ only beyond
leading order in slow roll (to leading order it is just given by
$Q_{mn}^\mathrm{lin} = c/(2 k^{3/2} R) \, \gd_{mn}$). Hence in a
leading-order slow-roll approximation the only non-linear parts in
the source terms are the factors in front of the integrals in
(\ref{sourcesSR}), plus the basis vector $e_p$ inside the $\ga_p$.

Within the leading-order slow-roll approximation, we now look at
the two-field case, to make things a bit more explicit. In that
case the matrix $A$ in (\ref{basic_sym}) is given by
    \beq\label{2fA}
    A = \pmtrx{0&-1&0&0\\0&3&-6\tget^\perp&0\\0&0&0&-1\\0&0&3\gc&3},
    \qquad\qquad
    \gc \equiv \frac{V_{22}}{3H^2} + \tge + \tget^\parallel.
    \eeq
The quantity $\gc$ is first order in slow roll. Here we used
(\ref{basicSR}) and the relation, valid to leading order in slow
roll (see e.g.\ \cite{vantent}),
    \beq
    \frac{V_{1m}}{3 H^2} = \frac{V_{m1}}{3 H^2}
    = (\tge - \tget^\parallel) \gd_{m1} - \tget^\perp \gd_{m2}.
    \eeq
Using the constraints (\ref{constr1})--(\ref{constr3}) we can
compute the spatial derivatives that are needed to calculate
$\bA$. Some of these were given in a general form in
\cite{formalism}; to first order in slow roll for the $\gth$
coefficients and to second order for the $\gz$ coefficients we
find in the two-field case,
    \bea
    (\der_i \ln a)^{(1)} & = & - (\der_i \ln H)^{(1)}
    = - \tge \, \gz_i^{(1)\,1} ,
    \non\\
    (\der_i \ln\tge)^{(1)} & = & - 2 \gth_i^{(1)\,1}
    - 2 (\tge+\tget^\parallel) \gz_i^{(1)\,1} + 2 \tget^\perp \gz_i^{(1)\,2},
    \label{deribasic}\\
    (\der_i \tget^\perp)^{(1)} & = & \tget^\perp \gth_i^{(1)\,1}
    + (3-3\tge+\tget^\parallel) \gth_i^{(1)\,2}
    - \lh (\tge-2\tget^\parallel)\tget^\perp + \tgx^\perp \rh \gz_i^{(1)\,1}
    + \lh 3\gc + (\tge+\tget^\parallel)\tget^\parallel - (\tget^\perp)^2 \rh
    \gz_i^{(1)\,2} ,
    \non\\
    (\der_i \gc)^{(1)} & = & (3-5\tge+\tget^\parallel) \gth_i^{(1)\,1}
    - 3\tget^\perp \gth_i^{(1)\,2} - \gps_1 \, \gz_i^{(1)\,1}
    - \lh \gps_2 + (6-10\tge+2\tget^\parallel+3\gc)\tget^\perp \rh
    \gz_i^{(1)\,2} ,
    \non
    \eea
introducing the two second-order slow-roll quantities $\gps_1$ and
$\gps_2$ as short-hand notation:
    \bea
    \gps_1 & \equiv & 2\tge\gc + (\tge-\tget^\parallel)
    \tget^\parallel + 3(\tget^\perp)^2 + \tgx^\parallel
    + \frac{\sqrt{2\tge}}{\gk} \frac{V_{221}}{3 H^2}
    \: = \: 2\tge (\gc + 2 \tget^\parallel) + 4(\tget^\perp)^2
    - \frac{\sqrt{2\tge}}{\gk} \frac{1}{3 H^2} \lh V_{111} - V_{221} \rh,
    \non\\
    \gps_2 & \equiv & (11\tge+2\tget^\parallel-3\gc)\tget^\perp
    +\tgx^\perp + \frac{\sqrt{2\tge}}{\gk} \frac{V_{222}}{3 H^2},
    \label{psi12}
    \eea
with $V_{mnp} \equiv e_m^A e_n^B e_p^C \cD_C \cD_B \der_A V$. The
reason for this specific definition of $\gps_2$ will become clear
later on. Since we have only two fields, the notation $\tgx^\perp$
is unambiguous. To compute $\der_i \gc$ we used that in the
two-field case, because of the orthonormality of the basis
vectors, $\cD_i e_2^A = - e_1^A (e^{\;}_{2\,B} \cD_i e_1^B)$. All
slow-roll parameters in these expressions take their homogeneous
background values. From this we find that the rank-3 matrix
$\bA_{abc}$ (defined below (\ref{tXdef})) is
    \beq\label{bA}
    \bA = \pmtrx{\boldsymbol{0}&\boldsymbol{0}&\boldsymbol{0}&\boldsymbol{0}\\
    \boldsymbol{0}&\boldsymbol{0}&
    -6 \lh -(\tge-2\tget^\parallel)\tget^\perp-\tgx^\perp,\tget^\perp,
    3\gc+(\tge+\tget^\parallel)\tget^\parallel-(\tget^\perp)^2,
    3-3\tge+\tget^\parallel \rh
    &\boldsymbol{0}\\\boldsymbol{0}&\boldsymbol{0}&\boldsymbol{0}
    &\boldsymbol{0}\\
    \boldsymbol{0}&\boldsymbol{0}&3 \lh -\gps_1,3-5\tge+\tget^\parallel,
    -\gps_2 - (6-10\tge+2\tget^\parallel+3\gc)\tget^\perp,-3\tget^\perp \rh
    &\boldsymbol{0}}.
    \eeq
In the same way we find that the matrices $X_{am}$ and
$\bX_{amc}$, defined in (\ref{tXdef}), are given by
    \bea\label{XbX}
    X & = & - \frac{\gk}{a \sqrt{2\tge}}
    \pmtrx{1&0\\0&2\tget^\perp\\0&1\\0&-\gc}
    \frac{\mathrm{e}^t}{2 k^{3/2}},
    \qquad
    \bX = - \frac{\gk}{a \sqrt{2\tge}}
    \pmtrx{(2\tge+\tget^\parallel,1,-\tget^\perp,0)
    &-(\tget^\perp,0,\tge+\tget^\parallel,1)\\
    2\tget^\perp(\tget^\perp,0,\tge+\tget^\parallel,1)&\vc{X}_{22}\\
    (\tget^\perp,0,\tge+\tget^\parallel,1)
    &(2\tge+\tget^\parallel,1,-\tget^\perp,0)\\
    -\gc(\tget^\perp,0,\tge+\tget^\parallel,1)&\vc{X}_{42}}
    \frac{\mathrm{e}^t}{2 k^{3/2}},
    \\
    \vc{X}_{22} & \equiv &
    2 \lh (\tge+3\tget^\parallel)\tget^\perp-\tgx^\perp,2\tget^\perp,
    3\gc+(\tge+\tget^\parallel)\tget^\parallel-2(\tget^\perp)^2,
    3-3\tge+\tget^\parallel \rh,
    \non\\
    \vc{X}_{42} & \equiv &
    \lh \gps_1-(2\tge+\tget^\parallel)\gc,-3+5\tge-\tget^\parallel-\gc,
    \gps_2+(6-10\tge+2\tget^\parallel+4\gc)\tget^\perp,3\tget^\perp \rh,
    \non
    \eea
where we used (\ref{linsol}), (\ref{sourcesSR}), and (\ref{deribasic}). Note
that $-(\der_i \ga_1)/\ga_2 = (\der_i \ga_2)/\ga_1 = \tget^\perp \gz_i^1
+ (\tge+\tget^\parallel)\gz_i^2 + \gth_i^2$ to leading order in slow roll.

In general the Green's function $G_{ab}(t,t')$ cannot be expressed
in closed form, since the time-dependent matrix $A$ does not
commute at different times. It can be formally represented as
    \beq
    G(t,t')=\cT\exp{\left[-\int^t_{t'}A(s)\d s\right]},
    \eeq
where $\cT$ denotes a time-ordered exponential:
    \beq\label{time_ordered}
    \cT\exp{\left[-\int^t_{t'}A(s)ds\right]}\equiv \bm{1} -\int^t_{t'}A(s)\d s
    +\int^t_{t'}\d s \int^s_{t'}\d s' A(s) A(s')
    -\int^t_{t'}\d s \int^s_{t'}\d s' \int^{s'}_{t'} \d s''
    A(s) A(s') A(s'') + \ldots\,\,.
    \eeq
This formal expression is standard in quantum mechanics and
quantum field theory (see e.g.\ \cite{P&S}) where, viewed as a
perturbative expansion, the first few terms in the series are kept
when the operator $A$ contains a small parameter. In our case,
however, not all elements of $A$ are first order in slow roll, so
that a truncation at any finite order is a bad approximation.
Moreover, even if $A$ were first order in slow roll, one should
still be careful, because the time interval in the integrations
can easily be of the order of an inverse slow-roll parameter. The
time-ordered exponential can be written as an ordinary exponential
plus terms which contain (nested) commutators. For example, the
second and third order terms in the series (\ref{time_ordered})
can be written as
    \beq
    \int^t_{t'}\d s \int^s_{t'}\d s' A(s) A(s') =
    \frac{1}{2}\left(\int^t_{t'}A(s)\d s\right)^2
    +\frac{1}{2}\int_{t'}^t \d s \int_{t'}^t \d s'
    \left[A(s),A(s')\right]\Theta(s-s')
    \eeq
and
    \equa{
    \int^t_{t'}\d s \int^s_{t'}\d s' \int^{s'}_{t'} \d s''
    & A(s) A(s') A(s'') =
    \frac{1}{3!} \left(\int^t_{t'}A(s) ds\right)^3+\frac{1}{2}
    \int_{t'}^t \d s \int_{t'}^t \d s' \int_{t'}^t \d s''
    A(s)\left[A(s'),A(s'')\right]\Theta(s'-s'')
    \nonumber\\
    & +\frac{1}{3}
    \int_{t'}^t \d s \int_{t'}^t \d s' \int_{t'}^t \d s''
    \Bigl(\left[A(s),A(s')\right]A(s'')-A(s'')\left[A(s),A(s')\right]
    \Bigr)\Theta(s-s')\Theta(s-s'')
    }
respectively, where $\gTh$ is the step function, and similarly for
higher orders (see \cite{lam} for general expressions).

There are basically three ways to proceed with this expression. In
the first place we can, if we are interested only in relatively
short time intervals, neglect the commutator terms in the
expansion of the time-ordered exponential and write it as an
ordinary exponential. The commutator terms all contain a
difference of slow-roll parameters at different times, as opposed
to the terms of the ordinary exponential that have just a
slow-roll parameter at one time. Hence, for small time intervals,
the commutator terms are a slow-roll order of magnitude smaller.
Then we have an exact analytic solution in closed form for the
Green's function. Secondly, as will be the case in the explicit
example in the next section, we can consider examples where $A$
does commute with itself at different times, in which case the
time-ordered exponential simplifies to an ordinary exponential
exactly. Finally, we can compute the Green's function numerically
and use it in a semi-analytic calculation (remember that the
Green's function has to be computed only once). That will be
worked out in a future publication, though we give some results in
section~\ref{realmodel}.

\section{Explicit solution for two-field slow-roll case}
\label{expl2fsec}

In this section we provide an analytic solution for the bispectrum
in two-field slow-roll inflation. We assume slow roll as in
section~\ref{SRapproxsec} but in order to obtain explicit solutions in
sections~\ref{A}, \ref{B} and \ref{C} we make the
\emph{further} assumption that all slow-roll parameters are
constant. The semi-analytic results of section~\ref{realmodel} are
\emph{not} bound by this further assumption and all slow-roll
parameters are calculated numerically for a quadratic model. No
slow-roll parameter takes values greater than unity. However, a
semi-analytic approach is feasible for the most general
non-slow-roll case and will be the study of a future publication
\cite{RSvTnum}.

\subsection{Power spectrum}
\label{A}

We now restrict ourselves to just two fields. Moreover, we assume
the background values of $H$ and all slow-roll parameters,
including perpendicular ones, to be completely constant in time
whenever they are the leading-order (in slow roll) non-zero terms
in our expressions. Then we can actually solve the system
explicitly, i.e.\ do the time integrals. We start from the
equation of motion (\ref{basic_sym}) for $v_i$ together with the
definitions (\ref{2fA}), or rather from (\ref{vimeq}) for the
$m$th order $v_i^{(m)}$ in an expansion in perturbation orders. We
assume everywhere that $\gc > 0$. At first order in the
perturbations this reads as
    \beq\label{eqv1}
    \dot{v}_i^{(1)} + A^{(0)} v_i^{(1)} = b_i^{(1)},
    \qquad\qquad
    \lim_{t\rightarrow-\infty} v_i^{(1)} = 0.
    \eeq
The matrix $A^{(0)}$ contains just background quantities, which by
assumption are constant. Hence we circumvent the issues of
non-commutativity and time-ordered exponentials, and we can write
down the solution immediately as
    \beq\label{solv1}
    v_i^{(1)}(t,\vc{x}) = \mathrm{e}^{-A^{(0)}t} \int_{-\infty}^t \d t'
    \mathrm{e}^{A^{(0)}t'} b_i^{(1)}(t',\vc{x}).
    \eeq
(In the terminology of section~\ref{generalsec}, the Green's
function is $G(t,t') = \exp[-A^{(0)}(t-t')]$.) The exponential can
be worked out using its eigenvalues and eigenvectors. When
multiplied with its inverse (at a different time) and expanded to
first order in slow roll, we obtain
    \equa{\label{expAt}
    & \mathrm{e}^{-A^{(0)}t} \mathrm{e}^{A^{(0)}t'} =
    \\
    & \quad
    \pmtrx{1 & \frac{1}{3} \lh 1 - (y/y')^3 \rh & \frac{2\tget^\perp}{\gc}
    \lh 1 - (1+\frac{\gc}{3})(y/y')^\gc + \frac{\gc}{3} (y/y')^{3-\gc} \rh
    & \frac{2\tget^\perp}{3\gc} \lh 1 - (y/y')^3 - (1+\frac{2\gc}{3})
    \lh (y/y')^\gc - (y/y')^{3-\gc} \rh \rh\\
    0 & (y/y')^3 & 2\tget^\perp \lh (y/y')^\gc - (y/y')^{3-\gc} \rh
    & \frac{2\tget^\perp}{\gc} \lh (y/y')^3 + \frac{\gc}{3} (y/y')^\gc
    - (1+\frac{\gc}{3}) (y/y')^{3-\gc} \rh\\
    0 & 0 & (1+\frac{\gc}{3}) (y/y')^\gc - \frac{\gc}{3} (y/y')^{3-\gc}
    & \frac{1}{3} (1+\frac{2\gc}{3}) \lh (y/y')^\gc - (y/y')^{3-\gc} \rh\\
    0 & 0 & -\gc \lh (y/y')^\gc - (y/y')^{3-\gc} \rh
    & -\frac{\gc}{3} (y/y')^\gc + (1+\frac{\gc}{3}) (y/y')^{3-\gc}}
    \non
    }
Obviously, it is the identity matrix if $y'=y$. For calculational
simplicity here we have defined the new time variable $y$, as well
as the relative momentum $p$,
    \beq\label{timeredef}
    y \equiv \frac{k_* c}{\sqrt{2}} \, \mathrm{e}^{-t} = \frac{k_* R}{\sqrt{2}}
    = \frac{c}{\sqrt{2}} \, \mathrm{e}^{-\gD t_*},
    \qquad\qquad
    p \equiv \frac{k}{k_*}
    \qquad\Rightarrow\qquad
    p y = \frac{kR}{\sqrt{2}} = \frac{c}{\sqrt{2}} \, \mathrm{e}^{-\gD t_k},
    \eeq
with $\gD t_* \equiv t - t_*$, the time since horizon crossing of
a reference mode $k_*$, and we have used the fact that $k_*
\exp(-t_*) = 1$ by definition ($\gD t_k$ is defined similarly for
the mode $k$). The fixed reference mode $k_*$ is most conveniently
chosen to be one of the observable modes, say the one that crossed
the horizon 50 e-folds before the end of inflation. From
(\ref{bia2}) and (\ref{XbX}) and the relation $a =
kc/(pyH\sqrt{2})$ we find that $b_i^{(1)}$ can be written to
leading order in slow roll as
    \beq\label{bi1}
    b_i^{(1)} = - \frac{\gk}{2\sqrt{2}} \frac{H}{\sqrt{\tge}} \dk
    \frac{1}{k^{3/2}} \, 2 p^2 y^2 \mathrm{e}^{-p^2 y^2} \,
    \mathrm{i}k_i \, \mathrm{e}^{\mathrm{i} \vc{k}\cdot\vc{x}}
    \pmtrx{\ga_1(\vc{k})\\2\tget^\perp \ga_2(\vc{k})\\\ga_2(\vc{k})\\
    -\gc \ga_2(\vc{k})}
    +\mathrm{c.c.}
    \eeq
Changing to $y$ as integration variable, we can then do the
integral in (\ref{solv1}) explicitly to find the solution
    \bea
    v_i^{(1)}(y,\vc{x}) & = & - \frac{\gk}{2\sqrt{2}} \frac{H}{\sqrt{\tge}} \dk
    \frac{1}{k^{3/2}} \, \mathrm{i}k_i \,
    \mathrm{e}^{\mathrm{i} \vc{k}\cdot\vc{x}}
    \pmtrx{\ga_1 + 2 \frac{\tget^\perp}{\gc} \, \ga_2 & 0 
    & - 2 \frac{\tget^\perp}{\gc}\,\ga_2 & 0\\
    0 & 0 & 2 \tget^\perp \ga_2 & 0\\
    0 & 0 & \ga_2 & 0\\
    0 & 0 & - \gc \ga_2 & 0}
    \pmtrx{\mathrm{e}^{-p^2 y^2}\\ p^3 y^3 \,
    \gG\lh -\frac{1}{2},p^2 y^2\rh\\
    p^\gc y^\gc \, \gG\lh 1-\frac{\gc}{2},p^2 y^2 \rh\\
    p^{3-\gc} y^{3-\gc} \, \gG\lh -\frac{1}{2} + \frac{\gc}{2},p^2 y^2 \rh}
    +\mathrm{c.c.}
    \non\\
    & \equiv & - \frac{\gk}{2\sqrt{2}} \frac{H}{\sqrt{\tge}} \dk
    \frac{1}{k^{3/2}} \, \mathrm{i}k_i \,
    \mathrm{e}^{\mathrm{i} \vc{k}\cdot\vc{x}}
    \, B(\vc{k}) \, u(py) +\mathrm{c.c.}
    \label{solv1expl}
    \eea
where we have omitted the explicit $\vc{k}$ dependence of the
$\ga$'s and the final expression defines the matrix $B(\vc{k})$
and vector $u(py)$.

It is interesting to look at the time behaviour of
(\ref{solv1expl}) in more detail. Note that in our leading-order
slow-roll approximation, differences $\gD t$ in the time variable
defined in (\ref{time}) are equal to differences in the number of
e-folds. A few e-folds\footnote{For example, 3 e-folds
is good enough if $c=3$ and $\gc=0.05$, and this result depends
only weakly on the values of $c$ and $\gc$.} after horizon
crossing the vector $u(py)$ in (\ref{solv1expl}) can be
approximated by $(1,0,p^\gc y^\gc \, \gG(1-\gc/2),0)^T$. The third
entry can be approximated even further as just $1-\gc\gD t_k$,
where $\gD t_k$ is the number of e-folds after horizon crossing of
the mode $k$, and the expression is valid for $\gc \gD t_k$
sufficiently smaller than unity, but $\gD t_k \simg 3$.\footnote{See
the previous footnote. A logarithmic dependence on $c$ has
been ignored here. For $c=3$ this term is 4 times smaller than
$\gc \gD t_k$ when $\gD t_k = 3$, and becomes even less important
as $\gD t_k$ grows.} With this approximation the solution
(\ref{solv1expl}) can be written as
    \beq
    v_i^{(1)}(t,\vc{x}) \approx - \frac{\gk}{2\sqrt{2}}
    \frac{H}{\sqrt{\tge}} \dk \frac{1}{k^{3/2}}\,\mathrm{i}k_i\,
    \mathrm{e}^{\mathrm{i} \vc{k}\cdot\vc{x}}
    \pmtrx{\ga_1 + 2 \tget^\perp \gD t_k \, \ga_2\\
    2 \tget^\perp (1-\gc\gD t_k)\ga_2\\
    (1-\gc\gD t_k)\ga_2\\
    - \gc(1-\gc\gD t_k)\ga_2}
    +\mathrm{c.c.}
    \label{solv1approx}
    \eeq
As expected (see e.g.\ \cite{vantent}) we find that the
effectively single-field ($\ga_1$) component of $\gz_i^1$ reaches
its constant final value right after horizon crossing, while the
influence of the perpendicular field direction ($\ga_2$,
`isocurvature mode') on $\gz_i^1$ continues to grow with time on
super-horizon scales. The velocities $\gth_i^1$ and $\gth_i^2$ are
both suppressed by an additional slow-roll factor compared to the
$\gz_i$'s. In the limit of $\gD t_k \rightarrow \infty$ (or $py
\rightarrow 0$), where the above approximation is no longer valid,
the exact result (\ref{solv1expl}) leads to the limit
    \beq
    v_i^{(1)}(\vc{x}) \approx - \frac{\gk}{2\sqrt{2}} \frac{H}{\sqrt{\tge}}
    \dk \frac{1}{k^{3/2}}\,\mathrm{i}k_i\,
    \mathrm{e}^{\mathrm{i} \vc{k}\cdot\vc{x}}
    \pmtrx{\ga_1 + 2 \frac{\tget^\perp}{\gc} \, \ga_2\\0\\0\\0}
    +\mathrm{c.c.}
    \label{solv1limit}
    \eeq
Hence the expression does not diverge as $t$ grows, but reaches a
well-defined value, which is independent of the smoothing
parameter $c$.

Concentrating now on the adiabatic ($e_1$) component of $\gz
\equiv \der^{-2} \der^i \gz_i$ we find from (\ref{solv1expl}) to
leading order in slow roll:
    \beq
    \gz^{(1)\,1}(t,\vc{x}) =
    - \frac{\gk}{2\sqrt{2}} \frac{H}{\sqrt{\tge}}
    \dk \frac{1}{k^{3/2}} \, \mathrm{e}^{\mathrm{i} \vc{k}\cdot\vc{x}}
    \left[ \mathrm{e}^{-p^2 y^2(\gD t_*)} \ga_1(\vc{k})
    + 2 \frac{\tget^\perp}{\gc} \,
    g(p,\gc,\gD t_*) \, \ga_2(\vc{k}) \right] + \mathrm{c.c.},
    \eeq
where $y$ as a function of $\gD t_*$ is given in
(\ref{timeredef}), $p=k/k_*$, and we have defined
    \beq\label{gdef}
    g(p,\gc,\gD t_*) \equiv \mathrm{e}^{-p^2 y^2} - p^\gc y^\gc \,
    \gG\lh 1-\frac{\gc}{2},p^2 y^2 \rh
    \approx 1 - p^\gc \mathrm{e}^{-\gc \gD t_*}
    = 1 - \mathrm{e}^{-\gc \gD t_k},
    \eeq
where the approximation is good from a few e-folds after horizon
crossing. Hence the two-point correlator is given by
    \beq
    \left\langle \gz^{(1)\,1}(t,\vc{x}) \, \gz^{(1)\,1}(t,\vc{x}')
    \right\rangle
    = \frac{\gk^2}{8} \frac{H^2}{\tge}
    \int \frac{\d^3 \vc{k}}{(2\pi)^3} \frac{1}{k^3}
    \left[ \mathrm{e}^{-2 p^2 y^2(\gD t_*)}
    + 4 \lh \frac{\tget^\perp}{\gc} \rh^2 g^2(p,\gc,\gD t_*) \right]
    \mathrm{e}^{\mathrm{i}\vc{k}\cdot(\vc{x}-\vc{x}')} +
    \mathrm{c.c.},
    \eeq
or, equivalently, for the power spectrum:
    \beq\label{powerspec2f}
    \left\langle \gz^{(1)\,1}(k,t) \gz^{(1)\,1}(k,t) \right\rangle
    = \frac{\gk^2}{4} \frac{H^2}{\tge} \frac{1}{k^3}
    \left[ \mathrm{e}^{-2 p^2 y^2(\gD t_*)}
    + 4 \lh \frac{\tget^\perp}{\gc} \rh^2 g^2(p,\gc,\gD t_*) \right] .
    \eeq
Here we used (\ref{Gcrn}) to take the average. Alternatively, we
could have used (\ref{powerspec}) directly. From a few e-folds
after horizon crossing, $\exp(-2 p^2 y^2) \approx 1$ and
$g(p,\gc,\gD t_*)$ is given by the final expression in
(\ref{gdef}), so that the power spectrum is independent of the
smoothing parameter $c$. Finally, we can compute the adiabatic
spectral index using the expressions in \cite{vantent2}, where the
$U_{P\,e}$ in that paper can be read off from (\ref{powerspec2f}),
once the transient horizon-crossing effects have disappeared, to
be $2(\tget^\perp/\gc)g(p,\gc,\gD t_*) e_2$,
    \beq\label{specind}
    n_\mathrm{ad}-1 = -4\tge - 2\tget^\parallel - 8\tget^\perp
    \frac{\tget^\perp}{\gc} \, g(p,\gc,\gD t_*) \,
    \frac{1-g(p,\gc,\gD t_*)}
    {1+4\lh\frac{\tget^\perp}{\gc}\rh^2 g^2(p,\gc,\gD t_*)}.
    \eeq

\subsection{Second-order solution}
\label{B}

At second order in the perturbations we expand all quantities in
$A$ and $b_i$ as explained in (\ref{expansion}), using
(\ref{deribasic}), resulting in the expressions in (\ref{bA}) and
(\ref{XbX}). Remember that superscripts within parentheses denote
the order in perturbation theory, while the superscripts without
parentheses indicate the component of the vector within the field
basis as defined in (\ref{Lmdef}). The resulting equation for
$v_i^{(2)}$ has the same structure as (\ref{eqv1}), but with a
different source term:
    \beq\label{eqv2}
    \dot{v}_i^{(2)} + A^{(0)} v_i^{(2)} = b_i^{(2)}
    - \pmtrx{0\\-6 \tget^{\perp\,(1)}\\0\\3 \gc^{(1)}} \gz_i^{(1)\,2},
    \qquad\qquad
    \lim_{t\rightarrow-\infty} v_i^{(2)} = 0,
    \eeq
where $b_i^{(2)}$ is the vector obtained by perturbing $H$, $\tge$, 
$\tget^\perp$, $\gc$, and $e_1$ and $e_2$ inside $\ga_1$ and $\ga_2$ 
in $b_i^{(1)}$ given in (\ref{bi1}). 
Explicitly, the right-hand side of equation
(\ref{eqv2}) is to leading order in slow roll given by
    \equa{\label{bi2}
    \frac{\gk^2}{8} \frac{H^2}{\tge} & \int\!\!\!\!\int \frac{\d^3 \vc{k} \,
    \d^3 \vc{k}'}{(2\pi)^3}
    \frac{\mathrm{i}k_i \, \mathrm{e}^{\mathrm{i} \vc{k}\cdot\vc{x}}}
    {k^\frac{3}{2} {k'}^\frac{3}{2}}
    \left[ 2 p^2 y^2 \mathrm{e}^{-p^2 y^2}
    \pmtrx{(2\tge+\tget^\parallel)\ga_1(\vc{k})-\tget^\perp\ga_2(\vc{k})
    &\ga_1(\vc{k})
    &-\tget^\perp\ga_1(\vc{k})-(\tge+\tget^\parallel)\ga_2(\vc{k})
    &-\ga_2(\vc{k})\\
    *&*&*&*\\
    \tget^\perp\ga_1(\vc{k})+(2\tge+\tget^\parallel)\ga_2(\vc{k})
    &\ga_2(\vc{k})
    &(\tge+\tget^\parallel)\ga_1(\vc{k})-\tget^\perp\ga_2(\vc{k})
    &\ga_1(\vc{k})\\
    *&*&*&*}
    \right. \non\\
    & \left. - 3 \pmtrx{0\\0\\1\\0}^T B(\vc{k}) u(py)
    \pmtrx{0&0&0&0\\2(\tge-2\tget^\parallel)\tget^\perp+2\tgx^\perp&
    -2\tget^\perp&-6\gc-2(\tge+\tget^\parallel)\tget^\parallel+2(\tget^\perp)^2
    &-6+6\tge-2\tget^\parallel\\0&0&0&0\\
    -\gps_1&3-5\tge+\tget^\parallel&
    -\gps_2-(6-10\tge+2\tget^\parallel+3\gc)\tget^\perp&-3\tget^\perp}
    \right]
    \non\\
    & \times
    \lh B(\vc{k}') u(qy) \mathrm{e}^{\mathrm{i} \vc{k}'\cdot\vc{x}}
    +\mathrm{c.c.} \rh + \mathrm{c.c.}
    \non\\
    \equiv \:
    \frac{\gk^2}{8} & \frac{H^2}{\tge} \int\!\!\!\!\int \frac{\d^3 \vc{k}
    \, \d^3 \vc{k}'}{(2\pi)^3}
    \frac{\mathrm{i}k_i \, \mathrm{e}^{\mathrm{i} \vc{k}\cdot\vc{x}}}
    {k^\frac{3}{2} {k'}^\frac{3}{2}}
    \left[ 2 p^2 y^2 \mathrm{e}^{-p^2 y^2} \tB(\vc{k})
    - 3 (0,0,1,0) B(\vc{k}) u(py) F \right]
    \lh B(\vc{k}') u(qy) \mathrm{e}^{\mathrm{i} \vc{k}'\cdot\vc{x}}
    +\mathrm{c.c.} \rh + \mathrm{c.c.}
    }
where we have defined $q \equiv k'/k_*$, as well as the matrices
$\tB(\vc{k})$ and $F$ in the last expression (the matrix $B$ and
vector $u$ were defined in (\ref{solv1expl})). The entries
indicated by an asterisk in the matrix $\tB$ are not given
explicitly here, but can be read off from (\ref{XbX}); they do not
contribute to $\gz_i^{(1)\,1}$ and $\gz_i^{(1)\,2}$ to leading
order in slow roll, because they are one order higher than the
corresponding entries in the first and third row, after
cancellations in the final result have been taken into account.
The solution for $v_i^{(2)}(t,\vc{x})$ is now given by the same
expression (\ref{solv1}) as $v_i^{(1)}(t,\vc{x})$, if one replaces
$b_i^{(1)}$ in that expression by (\ref{bi2}), though actually
calculating the time integral to obtain a completely explicit
expression is clearly more difficult.

To get all the time-dependent terms together, it is useful to
change from the matrix notation used above to a component
notation, as defined in (\ref{basic_sym}). We define the indices
$a,b,c,d,e,f$ running from 1 to 4 to label the components in the
4-dimensional $\{\gz_i^1,\gth_i^1,\gz_i^2,\gth_i^2\}$ space.
Moreover, we rewrite the matrix in (\ref{expAt}) as
    \beq
    \mathrm{e}^{-A^{(0)}t} \mathrm{e}^{A^{(0)}t'} = K_{abc} w_c(y,y'),
    \eeq
    \beq
    K \equiv \pmtrx{(1,0,0,0)&\frac{1}{3}(1,-1,0,0)&
    \frac{2\tget^\perp}{\gc}(1,0,-1-\frac{\gc}{3},\frac{\gc}{3})
    &\frac{2\tget^\perp}{3\gc}(1,-1,-1-\frac{2\gc}{3},1+\frac{2\gc}{3})\\
    \boldsymbol{0}&(0,1,0,0)&2\tget^\perp(0,0,1,-1)
    &\frac{2\tget^\perp}{\gc}(0,1,\frac{\gc}{3},-1-\frac{\gc}{3})\\
    \boldsymbol{0}&\boldsymbol{0}&(0,0,1+\frac{\gc}{3},-\frac{\gc}{3})
    &\frac{1}{3}(0,0,1+\frac{2\gc}{3},-1-\frac{2\gc}{3})\\
    \boldsymbol{0}&\boldsymbol{0}&\gc(0,0,-1,1)
    &(0,0,-\frac{\gc}{3},1+\frac{\gc}{3})},
    \qquad w(y,y') \equiv
    \pmtrx{1\\(y/y')^3\\(y/y')^\gc\\(y/y')^{3-\gc}},
    \non
    \eeq
which defines the rank-3 matrix $K$ and the vector $w(y,y')$. Then
the solution for $v_i^{(2)}$ can be written as
    \bea\label{via2sol}
    v_{i\,a}^{(2)}(y,\vc{x}) & = &
    \frac{\gk^2}{8} \frac{H^2}{\tge} \int\!\!\!\!\int \frac{\d^3 \vc{k} \,
    \d^3 \vc{k}'}{(2\pi)^3}
    \frac{\mathrm{i}k_i \, \mathrm{e}^{\mathrm{i} \vc{k}\cdot\vc{x}}}
    {k^\frac{3}{2} {k'}^\frac{3}{2}} K_{abc} \lh B_{de}(\vc{k}')
    \mathrm{e}^{\mathrm{i} \vc{k}'\cdot\vc{x}} +\mathrm{c.c.} \rh
    \\
    && \times \left[ \tB_{bd}(\vc{k})
    \int_y^\infty \d y' \, 2 p^2 y' \mathrm{e}^{-p^2 {y'}^2} w_c(y,y') u_e(qy')
    - 3 B_{3f}(\vc{k}) F_{bd}
    \int_y^\infty \frac{\d y'}{y'} \, w_c(y,y') u_e(qy') u_f(py')
    \right] +\mathrm{c.c.}
    \non
    \eea
(To be precise, $c$, $e$, and $f$ are actually indices in a
completely different space than the other indices, but it is also
4-dimensional.) It is useful to consider the contributions of the
two integral terms within the square brackets separately, since
they have a different origin (cf.\ (\ref{generalfkkreal})). The
first term is the variation of the stochastic source, represented
in (\ref{generalfkkreal}) by the $\bX$ term, and because of the
window function the integral only picks up a contribution around
horizon crossing (although this contribution is time-dependent
even later on, because of the dependence on $y$, not just $y'$, of
the Green's function). The second term is the variation of the
coefficients in the equation of motion, represented in
(\ref{generalfkkreal}) by the $\bA$ term, which is an integrated
effect up to the end of inflation, and is not present in
single-field inflation.  The leading-order coefficients in front
of the first term are first order in slow roll, while the ones in
front of the second term are second order\footnote{There are some
entries in  the product $F_{bd}B_{de}$ that are first order in
slow roll, but these exactly cancel when (\ref{via2sol}) is worked
out explicitly, so that the non-vanishing leading-order
coefficients are second order in slow roll.}, however, this can be
more than compensated by the larger integration interval.

In principle there are 80 different integrals here: 16 from the
first term and 64 from the second one. Some of the integrals can
be done analytically, but most have to be studied numerically.
However, of those 64 from the second term the only integrals that
matter are those that are secular, i.e.\ continue to grow with
time (up to a time of order $\gD t \sim \gc\inv$), since these
will be, roughly speaking, a slow-roll order of magnitude larger
at the end of inflation than the other integrals. Although we
studied all integrals more carefully, one can easily get an idea
of which integrals in the second term will be secular by looking
at the behaviour of the integrand for $y'\rightarrow 0$ (i.e.\ $t
\rightarrow\infty$): only the components with $e$ and $f$ either 1
or 3 are secular, since these are close to ${y'}\inv$ in that
limit. (Actually the components with $e$ and $f$ either 2 or 4 are zero 
in this slow-roll approximation, as can be seen from (\ref{solv1expl})). 
A slightly more careful analysis shows that, roughly speaking, 
the $c=2$ and $c=4$ components of those terms will be a
factor $\gc$ smaller (one gets a $3\inv$ instead of a $\gc\inv$
when integrating). Given that $B_{31}=0$, $f$ cannot be equal to
1, so in the end one expects the 4 integrals in the second term
with $c$ and $e$ both either 1 or 3 and $f=3$ to be dominant, and
that is confirmed by a careful numerical study. We denote the
$(c,e,f)=(1,1,3)$, $(1,3,3)$, $(3,3,3)$, and $(3,1,3)$ integrals
in the second term within the square brackets of (\ref{via2sol})
by $I_1(p,q,\gc,\gD t_*)$, $I_2(p,q,\gc,\gD t_*)$,
$I_3(p,q,\gc,\gD t_*)$, and $I_4(p,q,\gc,\gD t_*)$, respectively.

Regarding the 16 integrals of the first term the following can be
said. Because of the $y^3$ factor in front of the $c=2,4$
integrals, which cannot be completely canceled by factors coming
from the integral, these terms will become negligible after just a
few e-folds after horizon crossing. Of the remaining integrals
those with $e=2$ and $e=4$ are zero. Hence there are also only 4 distinct 
integrals here that have to be considered: those with $c=1,3$ and $e=1,3$
Again, these simple estimates are confirmed by careful
numerical study of the integrals. We denote the $(c,e)=(1,1)$,
$(1,3)$, $(3,3)$, and $(3,1)$ integrals in the
first term within the square brackets of (\ref{via2sol}) by
$J_1(p,q,\gD t_*)$, $J_2(p,q,\gc,\gD t_*)$, $J_3(p,q,\gc,\gD t_*)$,
and $J_4(p,q,\gc,\gD t_*)$, respectively.

Let us now investigate these 8 integrals. Half of them, viz.\
$I_3$, $I_4$, $J_3$, and $J_4$, are zero in the limit of
$t\rightarrow\infty$, but decrease slowly enough with time that
they should not be neglected at the end of inflation. Three of the
integrals can be done analytically:
    \bea
    J_1(p,q,\gD t_*) & = & \int_{y(\gD t_*)}^\infty \d y' \, 2 p^2 y'
    \mathrm{e}^{-(p^2+q^2){y'}^2}
    \: = \: \frac{p^2}{p^2+q^2} \, \mathrm{e}^{-(p^2+q^2)y^2},
    \non\\
    J_3(p,q,\gc,\gD t_*) & = & q^\gc y^\gc(\gD t_*)
    \int_{y(\gD t_*)}^\infty \d y' \, 2 p^2 y' \mathrm{e}^{-p^2{y'}^2}
    \gG \lh 1-\frac{\gc}{2}, q^2 {y'}^2 \rh
    \\
    & = & -\frac{q^2}{p^2+q^2} (p^2+q^2)^{\gc/2} \, y^\gc \,
    \gG \lh 1-\frac{\gc}{2},(p^2+q^2)y^2 \rh
    + q^\gc y^\gc \mathrm{e}^{-p^2 y^2}
    \gG \lh 1-\frac{\gc}{2},q^2 y^2 \rh,
    \non\\
    J_4(p,q,\gc,\gD t_*) & = & y^\gc(\gD t_*)
    \int_{y(\gD t_*)}^\infty \d y' \, 2 p^2 {y'}^{1-\gc}
    \mathrm{e}^{-(p^2+q^2){y'}^2}
    \: = \: \frac{p^2}{p^2+q^2} (p^2+q^2)^{\gc/2} \, y^\gc \,
    \gG \lh 1-\frac{\gc}{2},(p^2+q^2)y^2 \rh,
    \non
    \eea
where $y(\gD t_*)$ is given in (\ref{timeredef}). It is also
interesting to look at the behaviour of the integrals in the
limits of $p\rightarrow 0$ ($k\rightarrow 0$) and $q\rightarrow 0$
($k'\rightarrow 0$). For $p\rightarrow 0$ all integrals are zero.
For $q\rightarrow 0$ only $I_1$, $I_4$, $J_1$, and $J_4$ are
non-zero. The integrals $J_1$ and $J_4$ are given above, but in
this limit also $I_1$ and $I_4$ can be computed analytically (the
expression for $I_4$ is only valid from a few e-folds after
horizon crossing, i.e.\ for $py \ml 1$):
    \bea
    I_1(p,0,\gc,\gD t_*)
    & = & \int_{y(\gD t_*)}^\infty \d y' \, p^\gc {y'}^{-1+\gc} \,
    \gG \lh 1-\frac{\gc}{2},p^2{y'}^2 \rh
    \: = \: \frac{1}{\gc} \lh \mathrm{e}^{-p^2 y^2} - p^\gc y^\gc \,
    \gG \lh 1-\frac{\gc}{2},p^2 y^2 \rh \rh
    \: = \: \frac{1}{\gc}\,g(p,\gc,\gD t_*),
    \non\\
    I_4(p,0,\gc,\gD t_*)
    & = & p^\gc y^\gc(\gD t_*) \int_{y(\gD t_*)}^\infty \frac{\d y'}{y'} \,
    \gG \lh 1-\frac{\gc}{2},p^2{y'}^2 \rh
    \: \simeq \: - p^\gc y^\gc \ln(py) \,\, \gG \lh 1 - \frac{\gc}{2} \rh
    \: \approx \: (\gD t_* - \ln p) p^\gc \mathrm{e}^{-\gc \gD t_*},
    \eea
where $g(p,\gc,\gD t_*)$ is defined in (\ref{gdef}). Using the
results discussed in the text above equation (\ref{solv1approx}),
one sees that from a few e-folds after horizon crossing both start
growing linearly with $\gD t_k$ ($= \gD t_* - \ln p$), although
finally the limit $1/\gc$ is reached for $I_1$, while $I_4$ goes
to zero.

    \begin{figure}
    \subfigure[$\gc = 0.001$]{\includegraphics[width=8.5cm]{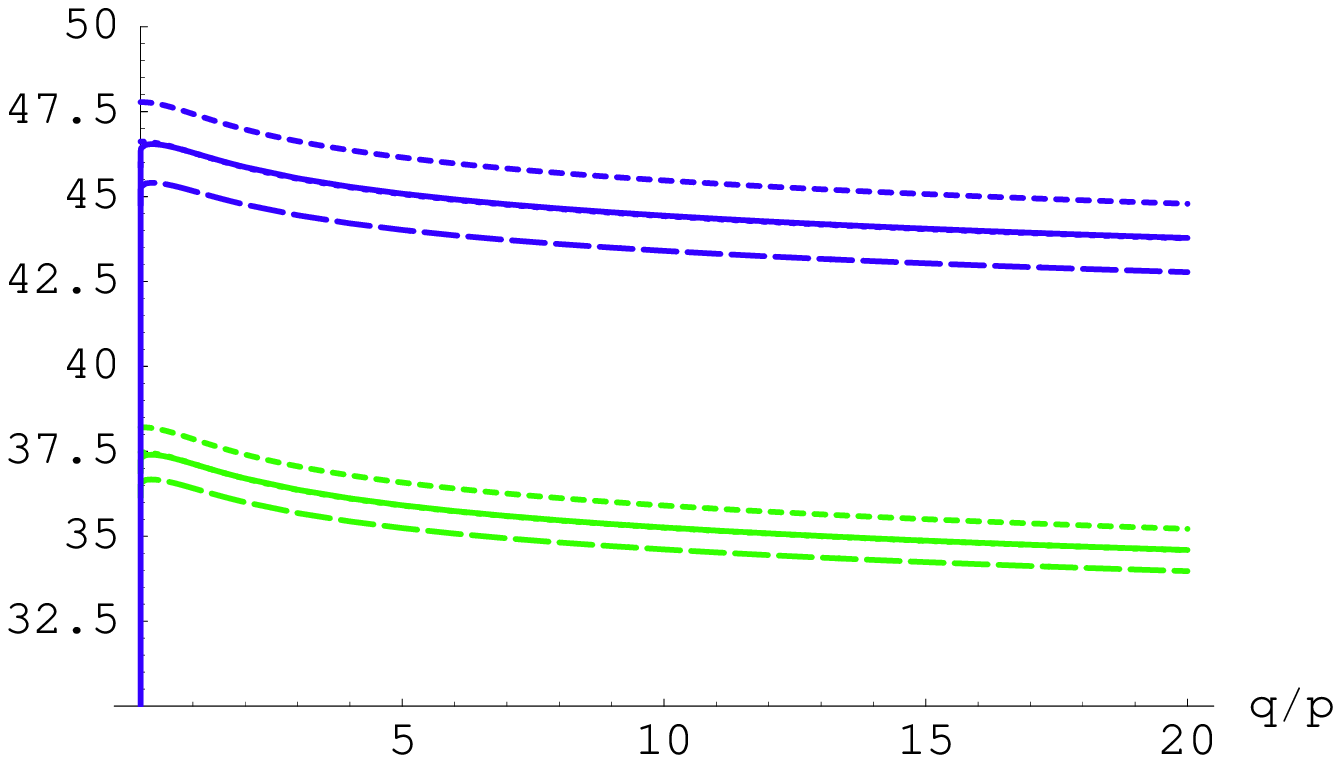}}
    \hspace{0.5cm}
    \subfigure[$\gc = 0.01$]{\includegraphics[width=8.5cm]{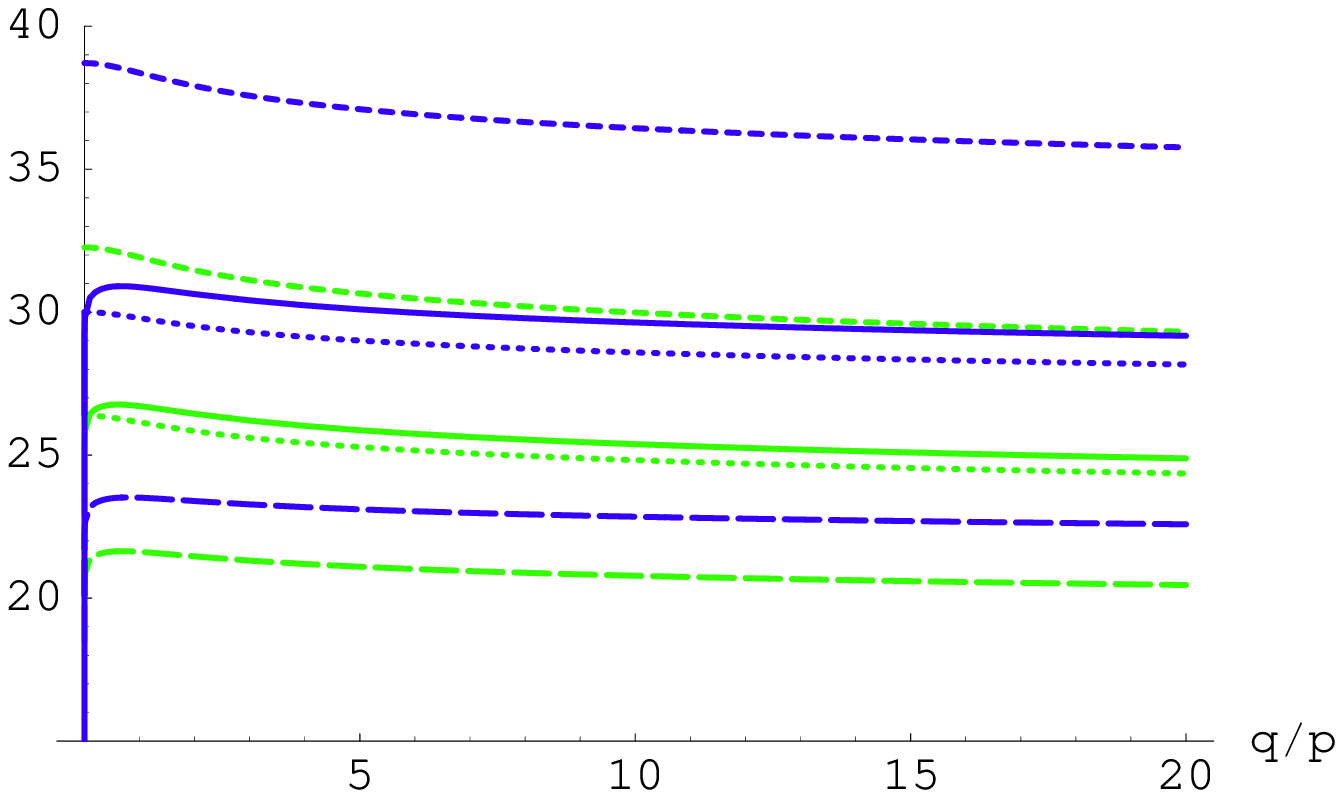}}\\
    \subfigure[$\gc = 0.05$]{\includegraphics[width=8.5cm]{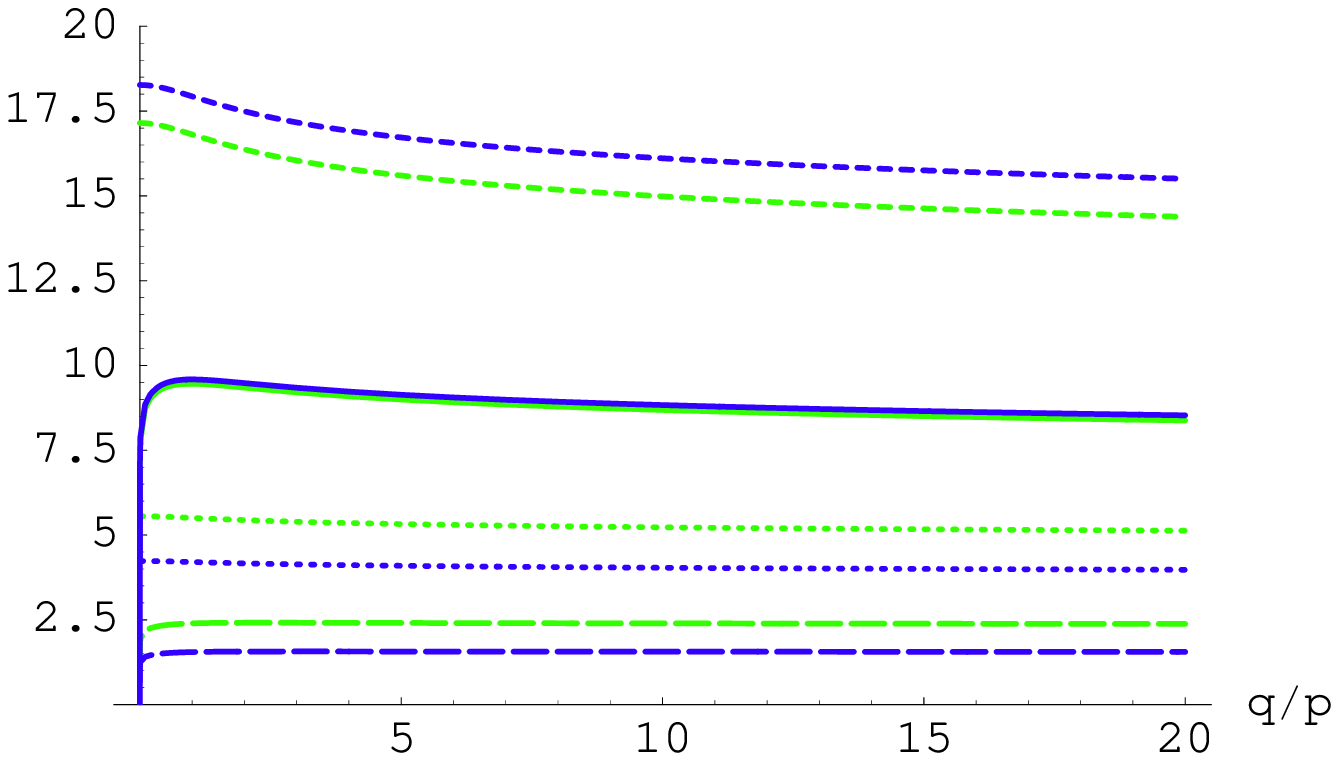}}
    \hspace{0.5cm}
    \subfigure[$\gc = 0.1$]{\includegraphics[width=8.5cm]{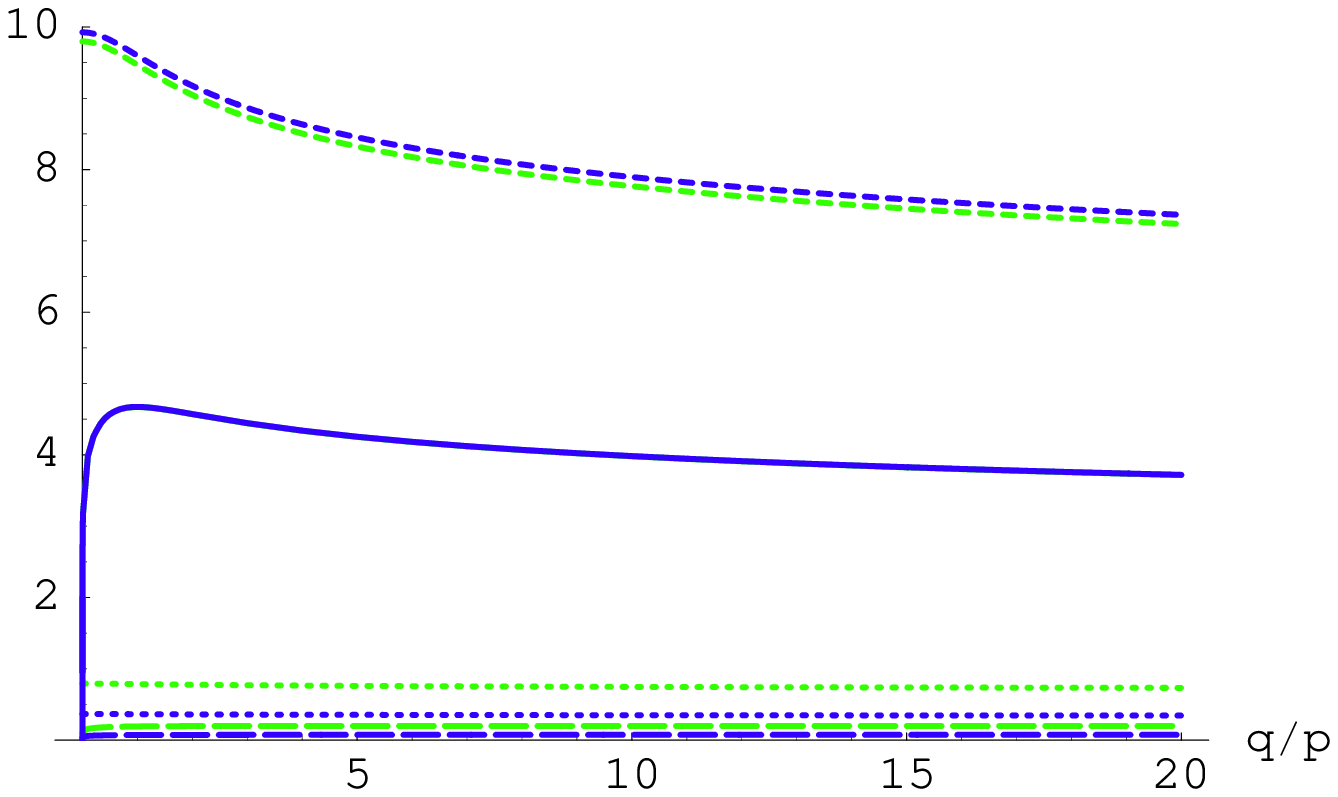}}
    \caption{The integrals $I_1(p,q,\gc,\gD t_*)$ (small dashes),
    $I_2(p,q,\gc,\gD t_*)$ (solid), $I_3(p,q,\gc,\gD t_*)$ (large dashes),
    and $I_4(p,q,\gc,\gD t_*)$ (dots) plotted as a function of $q/p$ for
    $\gD t_* = 50$, both for $p=1$ (blue, darker) and for $p=\exp(10)$
    (green, lighter) (i.e.\ 50 and 40 e-folds after horizon crossing of
    the mode $k$). The smoothing parameter $c=3$, although the dependence
    on $c$ is negligible. The different figures correspond with different
    values of $\gc$, as indicated. Note that $I_2$ and $I_4$ almost coincide
    in the first plot.}
    \label{integralfigs}
    \end{figure}

For $q>0$ the four $I$-integrals have to be evaluated numerically;
the resuls are plotted in figure~\ref{integralfigs} as a function
of $q/p$ for $\gD t_* = 50$, for various values of the parameters
$\gc$ and $p$. Although we will be using the exact numerical
results for all integrals when plotting the three-point
correlator, one can get an approximation by neglecting the $\gc/2$
inside the gamma function in $u_3(py')$ (see (\ref{via2sol}) and
(\ref{solv1expl})). Then the $I$-integrals can be done
analytically, with the results
    \equa{
    I_1(p,q,\gc,\gD t_*) & \approx \frac{1}{2} p^\gc (p^2+q^2)^{-\gc/2}
    \gG\lh \frac{\gc}{2},(p^2+q^2)y^2 \rh,
    & I_2(p,q,\gc,\gD t_*) & \approx \frac{1}{2} \, p^\gc q^\gc (p^2+q^2)^{-\gc}
    \gG\lh \gc,(p^2+q^2)y^2 \rh,
    \non\\
    I_3(p,q,\gc,\gD t_*) & \approx \frac{1}{2} \, p^\gc q^\gc
    (p^2+q^2)^{-\gc/2} y^\gc \gG\lh \frac{\gc}{2},(p^2+q^2)y^2 \rh,
    & I_4(p,q,\gc,\gD t_*) & \approx \frac{1}{2} \, p^\gc y^\gc
    \gG\lh 0,(p^2+q^2)y^2 \rh,
    }
so that we can make the following estimates:
    \equa{
    I_1 & \approx \gD t_k \lh 1-{\textstyle \frac{1}{2}}\gc\gD t_k
    +{\textstyle \frac{1}{6}}(\gc\gD t_k)^2 \rh,
    & I_2 & \approx \gD t_k \lh 1-\gc\gD t_k
    +{\textstyle \frac{2}{3}}(\gc\gD t_k)^2 \rh,
    \non\\
    I_3 & \approx \gD t_k \lh 1-{\textstyle \frac{3}{2}}\gc\gD t_k
    +{\textstyle \frac{7}{6}}(\gc\gD t_k)^2 \rh,
    & I_4 & \approx \gD t_k \lh 1-\gc\gD t_k
    +{\textstyle \frac{1}{2}}(\gc\gD t_k)^2 \rh,
    \label{intapprox1}
    }
for $\gc\inv \mg \gD t_k$, and
    \equa{
    I_1 & \approx \gc\inv,
    & I_2 & \approx {\textstyle \frac{1}{2}} \gc\inv,
    & I_3 & \approx 0,
    & I_4 & \approx 0,
    }
for $\gc\inv \ml \gD t_k$. 
As a rough approximation, they can be taken independent of $q$ for
reasonable ranges, say up to $q/p \sim 100$. For $I_2$ and $I_3$
this range has a lower limit as well: $100\inv \siml q/p \siml
100$; they are zero for $q=0$. Note that these secular
$I$-integrals typically give a result which is of the order of an
inverse slow-roll parameter.

    \begin{figure}
    \subfigure[\label{Jfigsa}]{\includegraphics[width=8.5cm]{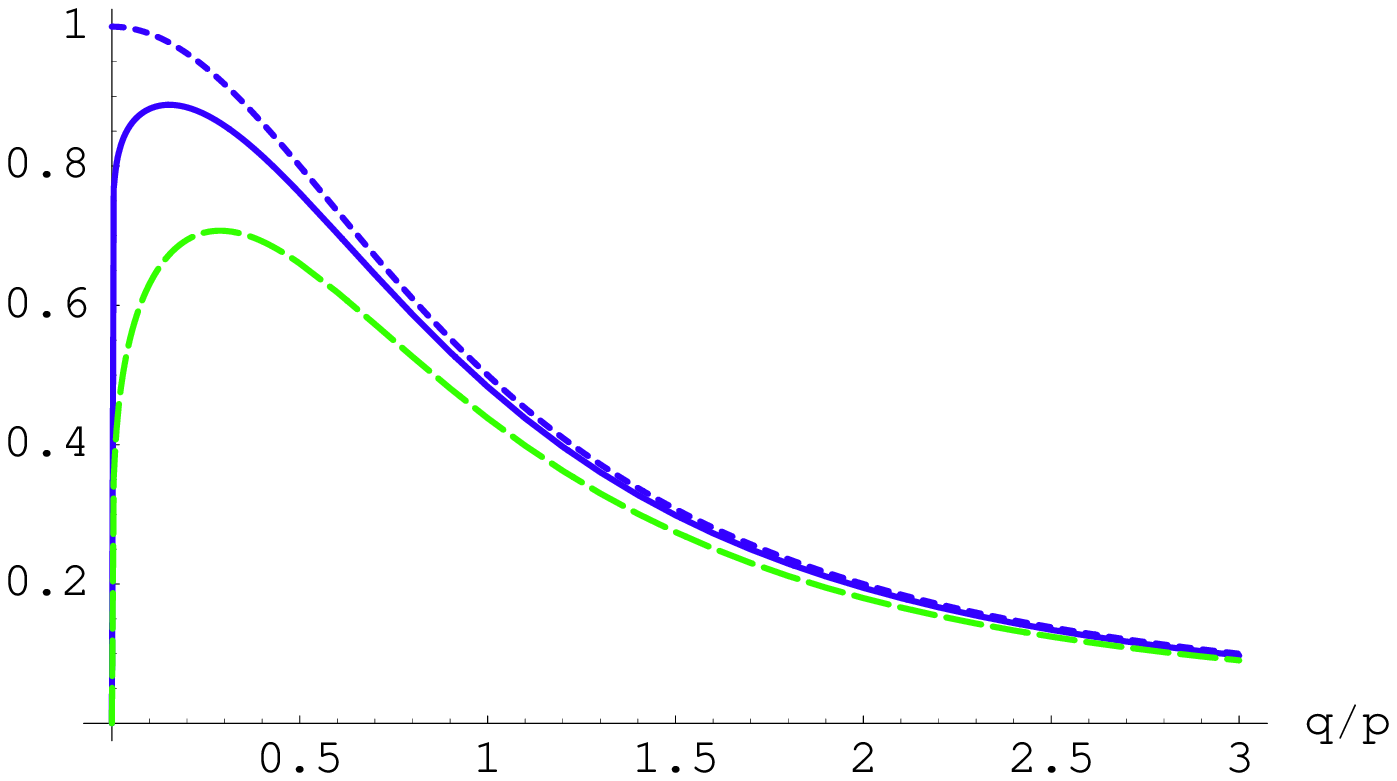}}
    \hspace{0.5cm}
    \subfigure[\label{Jfigsb}]{\includegraphics[width=8.5cm]{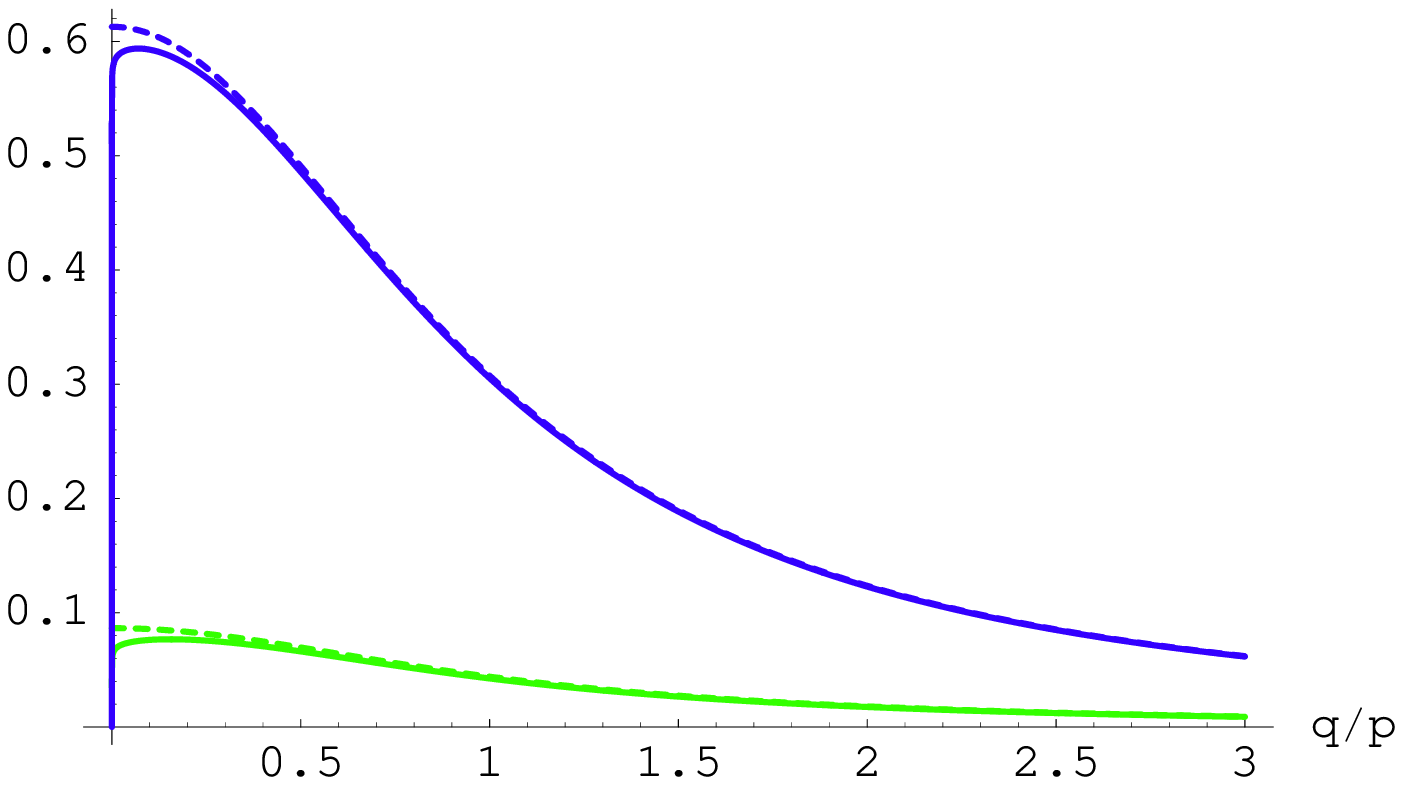}}
    \caption{(a) The integrals $J_1(p,q,\gD t_*)$ (small dashes)
    and $J_2(p,q,\gc,\gD t_*)$ for $\gc=0.05$ (solid)
    and $\gc=0.2$ (large dashes) plotted as a function of $q/p$,
    a few e-folds after horizon crossing of the mode $k$ when the
    dependence on $\gD t_k = \gD t_* - \ln p$ has become negligible.
    (b) The integrals $J_3(p,q,\gc,\gD t_*)$ (solid) and 
    $J_4(p,q,\gc,\gD t_*)$ (small dashes) plotted as a function of $q/p$ 
    for $p=1$ and $\gD t_* = 50$, for both
    $\gc=0.01$ (blue, darker) and $\gc=0.05$ (green, lighter).}
    \end{figure}

For the $J$-integrals the results are much smaller, since the
integration interval is restricted because of the window function.
The integrals $J_1$ and $J_2$ become completely
independent of $\gD t_k$ from a few e-folds after horizon crossing
of the mode $k$. Moreover, $J_1$ is independent of
$\gc$, while $J_2$ has a relatively weak dependence on $\gc$. On
the other hand, both depend strongly on $q/p$. They are
plotted in figure~\ref{Jfigsa}. The integrals $J_3$ and $J_4$ depend 
strongly on both $q/p$ and $\gc \gD t_k$. In the
limit $\gc\gD t_k \ml 1$ they become equal to $J_2$ and $J_1$, 
respectively, while in the opposite limit they both go to
zero. They are plotted in figure~\ref{Jfigsb}. For $q=0$ we have
an exact analytic result; for $q=p$ (i.e.\ $k=k'$) it is sometimes
useful to have an analytic approximation:
    \equa{
    J_1 & \approx {\textstyle \frac{1}{2}},
    & J_2 & \approx {\textstyle \frac{1}{2}},
    & J_3 & \approx {\textstyle \frac{1}{2}}(1-\gc\gD t_k),
    & J_4 & \approx {\textstyle \frac{1}{2}}(1-\gc\gD t_k),
    }
for $\gc\inv \mg \gD t_k$, and
    \equa{
    J_1 & \approx {\textstyle \frac{1}{2}},
    & J_2 & \approx {\textstyle \frac{1}{2}},
    & J_3 & \approx 0,
    & J_4 & \approx 0,
    \label{intJapprox2}
    }
for $\gc\inv \ml \gD t_k$.

Having studied all the integrals, we can now work out
(\ref{via2sol}) explicitly. We focus on the $a=1$ component of
$v_{i\,a}$, that is $\gz_i^1$ (the adiabatic component of
$\gz_i$), since that is the quantity we want to compute the
three-point correlator of in the end. The final result for
$\gz_i^{(2)\,1}$ in the two-field case, in a leading-order
slow-roll approximation (constant slow-roll parameters) and valid
well after horizon crossing, is
    \bea
    \gz_i^{(2)\,1}(t,\vc{x}) & = & \frac{\gk^2}{8} \frac{H^2}{\tge}
    \int\!\!\!\!\int \frac{\d^3 \vc{k} \, \d^3 \vc{k}'}{(2\pi)^3}
    \frac{1}{k^\frac{3}{2} {k'}^\frac{3}{2}}
    \, \mathrm{e}^{\mathrm{i} \vc{k}'\cdot\vc{x}}
    \label{gzi21sol}\\
    && \times
    \Biggl[ \lh \mathrm{i}k_i \, \mathrm{e}^{\mathrm{i} \vc{k}\cdot\vc{x}}
    \, \ga_1(\vc{k}) + \mathrm{c.c.} \rh
    \Biggl\{ \left[ (2\tge+\tget^\parallel) J_1 
    + \frac{2(\tget^\perp)^2}{\gc} (J_1-J_4) \right] \ga_1(\vc{k}')
    \non\\
    && \qquad\qquad 
    + 2 \frac{\tget^\perp}{\gc} \left[ (2\tge+\tget^\parallel) (J_1-J_3)
    - \tge (J_2-J_3) + \frac{2(\tget^\perp)^2}{\gc}(J_1-J_2+J_3-J_4) 
    - \frac{\gc}{2} (J_2-2J_3) \right] \ga_2(\vc{k}') \Biggr\}
    \non\\
    && \quad\;
    + 2 \frac{\tget^\perp}{\gc} \lh \mathrm{i}k_i \,
    \mathrm{e}^{\mathrm{i} \vc{k}\cdot\vc{x}} \, \ga_2(\vc{k})
    + \mathrm{c.c.} \rh
    \Biggl\{ \left[ (2\tge+\tget^\parallel) (J_1-J_4) - \frac{\gc}{2} \, J_1
    + \gps_1 (I_1-I_4) + \go_1 \, I_1 \right] \ga_1(\vc{k}')
    \non\\
    && \qquad\qquad
    + 2 \frac{\tget^\perp}{\gc} \Biggl[
    (2\tge+\tget^\parallel) (J_1-J_2+J_3-J_4)
    - \frac{\gc}{2} (J_1-2J_2+J_3) - \frac{\gc^2}{4(\tget^\perp)^2}
    (\tge+\tget^\parallel-\gc) J_2
    \non\\
    && \qquad\qquad
    + \gps_1 (I_1-I_2+I_3-I_4)
    + \go_1 (I_1-I_2) + \frac{\gc}{2\tget^\perp} \, \gps_2 (I_2-I_3)
    + \frac{\gc^2}{2(\tget^\perp)^2} \, \go_2 \, I_2
    \Biggr] \ga_2(\vc{k}')
    \Biggr\} \Biggr]
    + \mathrm{c.c.},
    \non
    \eea
with $\go_1 \equiv \gc(-\tge+2\tget^\parallel-\tgx^\perp/\tget^\perp)$
and $\go_2 \equiv (\tge+\tget^\parallel)\tget^\parallel +
(3\tge-\tget^\parallel)\gc + (\tget^\perp)^2$ defined as
short-hand notation. The arguments $(p,q,\gc,\gD t_*)$ of the
integrals have been suppressed, but it should of course be kept in
mind that that is where the time dependence resides in this
expression. Note that in the single-field limit, where all terms
with $\ga_2$ disappear and $\tget^\perp=0$, we recover the result of \cite{sf}. 
For the three-point correlator we need to know $\gz^{(2)\,1} \equiv
\der^{-2} \der^i \gz_i^{(2)\,1}$, which is given by the same
expression (\ref{gzi21sol}), but with $\mathrm{e}^{\mathrm{i}
\vc{k}'\cdot\vc{x}}\lh \mathrm{i}k_i \, \mathrm{e}^{\mathrm{i}
\vc{k}\cdot\vc{x}} \, \ga_1(\vc{k}) + \mathrm{c.c.} \rh$ replaced
by
    \beq
    \lh \frac{k^2 + \vc{k}\cdot\vc{k}'}{|\vc{k}+\vc{k}'|^2} \,
    \mathrm{e}^{\mathrm{i} (\vc{k}+\vc{k}')\cdot\vc{x}} \, \ga_1(\vc{k})
    + \frac{k^2 - \vc{k}\cdot\vc{k}'}{|\vc{k}-\vc{k}'|^2} \,
    \mathrm{e}^{-\mathrm{i} (\vc{k}-\vc{k}')\cdot\vc{x}} \, \ga_1^*(\vc{k}) \rh
    \eeq
and the same for $\ga_2(\vc{k})$.

\subsection{Bispectrum} 
\label{C}

As in the single-field case, $\langle \gz^{(2)\,1} \rangle$ is
indeterminate. To remove this ambiguity and also require that
perturbations have a zero average, we define $\tgz^m \equiv \gz^m
- \langle \gz^m \rangle$. Expanding $\tgz^m =
\tgz^{(1)\,m}+\tgz^{(2)\,m}$ and switching over to Fourier space,
we finally arrive at our end result for the three-point correlator
(or rather the bispectrum) of the adiabatic component:
    \beq\label{3pcorr}
    \left\langle \tgz^1(t,\vc{x}_1) \tgz^1(t,\vc{x}_2)
    \tgz^1(t,\vc{x}_3)
    \right\rangle^{(2)}(\vc{k}_1,\vc{k}_2,\vc{k}_3)
    = (2\pi)^3 \gd^3 (\vc{k}_1 + \vc{k}_2 + \vc{k}_3)
    \Bigl[ f(\vc{k}_1,\vc{k}_2) + f(\vc{k}_1,\vc{k}_3)
    + f(\vc{k}_2,\vc{k}_3) \Bigr]
    \eeq
with
    \bea\label{three point}
    f(\vc{k},\vc{k}') & \equiv & \frac{\gk^4}{16} \frac{1}{k^3 {k'}^3}
    \frac{H^4}{\tge^2} \frac{k^2 + \vc{k}\cdot\vc{k}'}{|\vc{k}+\vc{k}'|^2}
    \Biggl\{ (2\tge+\tget^\parallel) J_1
    + \frac{2(\tget^\perp)^2}{\gc} (J_1-J_4)
    \\
    && + 4 \lh \frac{\tget^\perp}{\gc} \rh^2 \Biggl[ 
    g(q,\gc,\gD t_*) \left[ (2\tge+\tget^\parallel) (J_1-J_3)
    - \tge (J_2-J_3) + \frac{2(\tget^\perp)^2}{\gc}(J_1-J_2+J_3-J_4) 
    - \frac{\gc}{2} (J_2-2J_3) \right] 
    \non\\
    && \qquad\qquad\qquad
    + g(p,\gc,\gD t_*)
    \left[ (2\tge+\tget^\parallel) (J_1-J_4) - \frac{\gc}{2} \, J_1
    + \gps_1 (I_1-I_4) + \go_1 \, I_1
    \right] \Biggr]
    \non\\
    && \hspace{-1cm} + 16 \lh \frac{\tget^\perp}{\gc} \rh^4 g(p,\gc,\gD t_*)
    g(q,\gc,\gD t_*)
    \Biggl[ (2\tge+\tget^\parallel) (J_1-J_2+J_3-J_4)
    - \frac{\gc}{2} (J_1-2J_2+J_3) 
    - \frac{\gc^2}{4(\tget^\perp)^2} (\tge+\tget^\parallel-\gc) J_2
    \non\\
    && \qquad\qquad\qquad
    + \gps_1 (I_1-I_2+I_3-I_4)
    + \go_1 (I_1-I_2)
    + \frac{\gc}{2\tget^\perp} \, \gps_2 (I_2-I_3)
    + \frac{\gc^2}{2(\tget^\perp)^2} \, \go_2 \, I_2
    \Biggr]
    \Biggr\} + \vc{k} \leftrightarrow \vc{k}' .
    \non
    \eea
Again, this result is valid in the two-field case, in a
leading-order slow-roll approximation (constant slow-roll
parameters) and valid from a sufficient number of e-folds after
horizon crossing that transient effects have disappeared. The
function $g(p,\gc,\gD t_*)$ is given in (\ref{gdef}), $\gc$,
$\gps_1$, and $\gps_2$ are defined in section~\ref{SRapproxsec},
and $\go_1$ and $\go_2$ are defined below (\ref{gzi21sol}). 
Remember that all the
integrals and the function $g(p,\gc,\gD t_*)$ depend on the
momenta via $p=k/k_*$ and $q=k'/k_*$ and hence are affected by the
interchange of $\vc{k}$ and $\vc{k}'$. In the single-field limit
only the first term on the first line of (\ref{three point}) remains, 
which agrees exactly with \cite{sf}.

\begin{figure}
    \includegraphics[width=14cm]{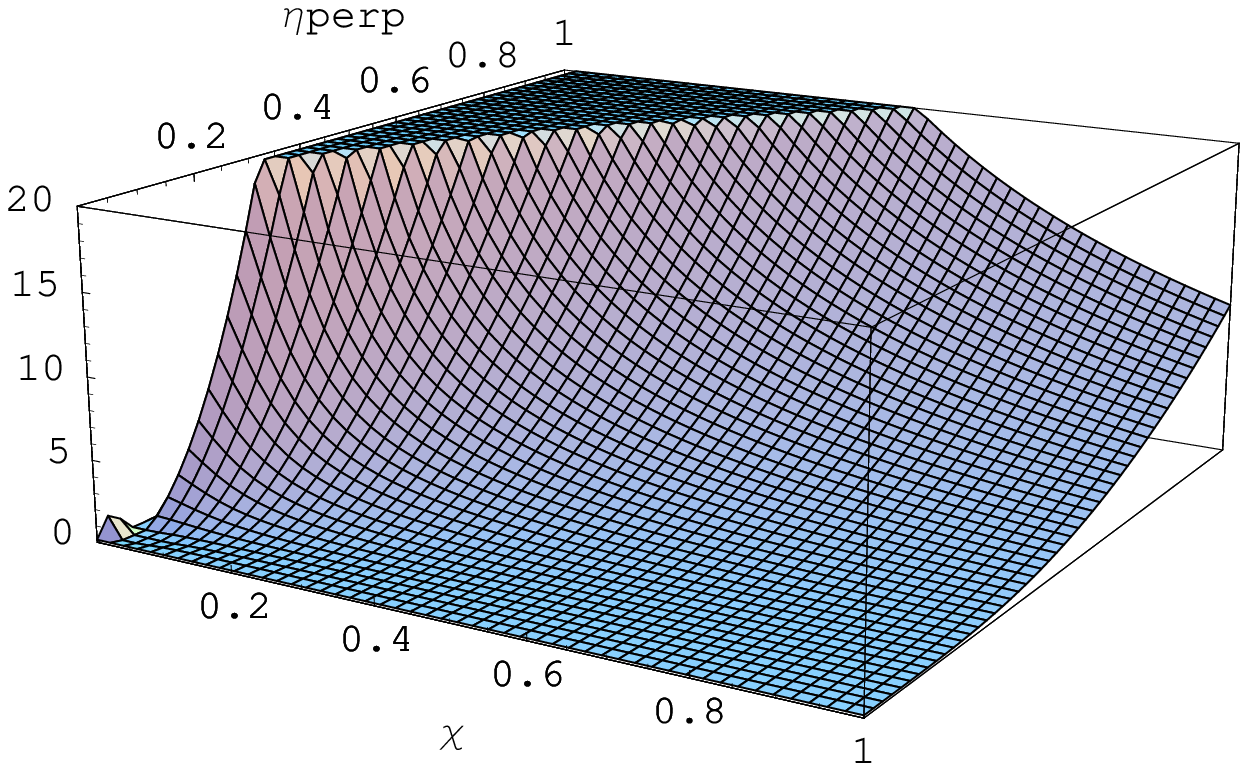}
    \caption{The bispectrum (\ref{3pcvertex}) divided by the square of the
    power spectrum (\ref{powerspec2f}) (with two different momenta)
    in the limit where one of the momenta is much smaller than the other two,
    plotted as a function of $\tget^\perp$ and $\gc$ for $\gD t = 50$,
    $\tge=-\tget^\parallel=0.05$, and $(\sqrt{2\tge}/\gk)V_{111}/(3 H^2)=
    -\tgx^\perp=-(\sqrt{2\tge}/\gk)V_{221}/(3 H^2)=0.003$.}
    \label{etaperpchifig}
    \end{figure}

In the limit $k_3 \ml k_1,k_2$ (and hence $\vc{k}_1 = - \vc{k}_2
\equiv \vc{k}$, while we also fix $\vc{k}_* = \vc{k}$ so that we
do not need to write a subscript on $\gD t$), all the integrals
can be performed analytically and the result is (leaving aside the
overall factor of $(2\pi)^3 \gd^3(\sum_s \vc{k}_s)$):
    \equa{\label{3pcvertex}
    \left\langle \tgz^1 \tgz^1 \tgz^1 \right\rangle^{(2)}
    = \: \frac{\gk^4}{8} \frac{1}{k^3 k_3^3} \frac{H^4}{\tge^2}
    \lh 1 + 4 \lh \frac{\tget^\perp}{\gc} \rh^2 \rh
    & \Biggl\{ (2\tge+\tget^\parallel)
    \lh 1 + 4 \lh \frac{\tget^\perp}{\gc} \rh^2
    \lh 1 - \mathrm{e}^{-\gc\gD t} \rh^2 \rh
    \\
    & \!\! + 4 \lh \frac{\tget^\perp}{\gc} \rh^2
    \lh 1 - \mathrm{e}^{-\gc\gD t} \rh
    \left[ \frac{\gps_1}{\gc} \lh 1 - (1+\gc\gD t)\mathrm{e}^{-\gc\gD t} \rh
    + \frac{\go_1}{\gc} \lh 1 - \mathrm{e}^{-\gc\gD t} \rh
    \right] \Biggr\} ,
    \non
    }
where the term on the first line within the curly brackets comes
from the $J$-integrals, and the term on the second line from the
$I$-integrals. Again, this agrees with the single-field result in
the limit $\tget^\perp \rightarrow 0$. Unlike the single-field
case, the multiple-field result cannot be expressed in terms of
the scalar spectral index and the power spectrum only (see
(\ref{powerspec2f}) and (\ref{specind}), and \cite{vantent2} for
expressions for the isocurvature and mixing components). Instead
of the three-point correlator itself, it is actually more useful
to look at the ratio of the bispectrum to the square of the power
spectrum, since that ratio is related to observables like the
$f_\mathrm{NL}$ parameter (more about that later). Dividing
(\ref{3pcvertex}) by the square of (\ref{powerspec2f}) (one with
momentum $k$ and the other with $k_3$) and taking the limit of
$\tget^\perp/\gc \mg 1$, we get for the two opposite limits of
$\gc \gD t$ that
    \beq\label{vertexratio}
    \frac{\langle \tgz^1 \tgz^1 \tgz^1 \rangle}
    {(\langle \tgz^1 \tgz^1 \rangle)^2}
    = \left\{ \begin{array}{ll} \displaystyle
    2 \lh \tge+3\tget^\parallel-\frac{\tgx^\perp}{\tget^\perp} \rh
    + \gps_1 \gD t
    & \mbox{for $\gc\inv \mg \gD t$},
    \\ \displaystyle
    2 \lh \tge+3\tget^\parallel-\frac{\tgx^\perp}{\tget^\perp}
    +\frac{\gps_1}{\gc} \rh
    & \mbox{for $\gc\inv \ml \gD t$}.
    \end{array} \right.
    \eeq
Now if we assume that $\tget^\perp$ is larger than the other
slow-roll parameters, the dominating term in both expressions will
be the $4(\tget^\perp)^2$ in $\gps_1$ (\ref{psi12}), so that the
two expressions in (\ref{vertexratio}) will go to
$4(\tget^\perp)^2 \gD t$ and $8(\tget^\perp)^2/\gc$, respectively.
Hence, while this ratio of the bispectrum to the square of the
power spectrum is first order in slow roll by naive power counting
(counting $1/\gD t$ as a slow-roll parameter), as in the
single-field case, it can be much larger for models with a
relatively small $\gc$ and relatively large $\tget^\perp$. For
example, $\tget^\perp=0.07$ and  $\gD t = 1/\gc = 50$ would
already give a ratio of more than unity, so that a value about 100
times larger than in the single-field case seems well within range
for multiple-field models. This is confirmed by the full plot of
(\ref{3pcvertex}) divided by the square of (\ref{powerspec2f}) as
a function of $\tget^\perp$ and $\gc$ given in
figure~\ref{etaperpchifig}. It is also interesting to see that in
the cases where non-Gaussianity is large, this is caused by the
$I$-integrals  (i.e.\ the super-horizon integrated background
effects that are absent in single-field inflation): roughly
speaking it boils down to $\tge J_1$ versus $(\tget^\perp)^2 I_1$,
which gives $\tge$ versus the smaller of $(\tget^\perp)^2 \gD t$
and $(\tget^\perp)^2/\gc$, either of which can easily be two
orders of magnitude larger.

In the opposite limit of $k_1=k_2=k_3 \equiv k$ (where we again
set $k_* = k$ so that $\gD t$ is unambiguous) we do not have an
exact analytic result for all integrals, but we can use the
approximations (\ref{intapprox1})--(\ref{intJapprox2}). We find
the following results for the bispectrum divided by the square of
the power spectrum, in the limit of $\tget^\perp/\gc \mg 1$:
    \beq\label{centreratio}
    \frac{\langle \tgz^1 \tgz^1 \tgz^1 \rangle}
    {(\langle \tgz^1 \tgz^1 \rangle)^2}
    = \left\{ \begin{array}{ll} \displaystyle
    \frac{3}{2} \lh \frac{1}{2 \gD t}
    + \frac{\go_1}{\gc}
    + \frac{\gps_2}{2\tget^\perp} + \frac{1}{3} \, \gps_1 \gD t \rh
    & \mbox{for $\gc\inv \mg \gD t$},
    \\ \displaystyle
    \frac{3}{2} \lh \frac{\gc}{2} 
    + \frac{\go_1}{\gc}
    + \frac{\gps_2}{2 \tget^\perp} + \frac{\gps_1}{\gc} \rh
    & \mbox{for $\gc\inv \ml \gD t$}.
    \end{array} \right.
    \eeq
If we assume once again that $\tget^\perp$ is larger than the
other slow-roll parameters, the dominating term in both
expressions will again be the $4(\tget^\perp)^2$ in $\gps_1$, so
that the two expressions will go to $2(\tget^\perp)^2 \gD t$ and
$6(\tget^\perp)^2/\gc$, respectively. Finally we should check that
all the limits that produce large non-Gaussianity do not produce
an unacceptably large spectral index at the same time. Fortunately
that is not the case: from (\ref{specind}) we derive, under the
same limits as in (\ref{vertexratio}) and (\ref{centreratio}),
    \beq\label{spec}
    n_\mathrm{ad} - 1
    = \left\{ \begin{array}{ll} \displaystyle
    -4 \tge - 2 \tget^\parallel - \frac{8(\tet^\perp)^2\Delta t}
    {1+4(\tet^\perp)^2(\Delta t)^2}
    & \mbox{for $\gc\inv \mg \gD t$},
    \\ \displaystyle
    -4 \tge - 2 \tget^\parallel
    & \mbox{for $\gc\inv \ml \gD t$}.
    \end{array} \right.
    \eeq

\begin{figure}
    \subfigure[\label{3pcorrfig}]
    {\includegraphics[width=11.5cm]{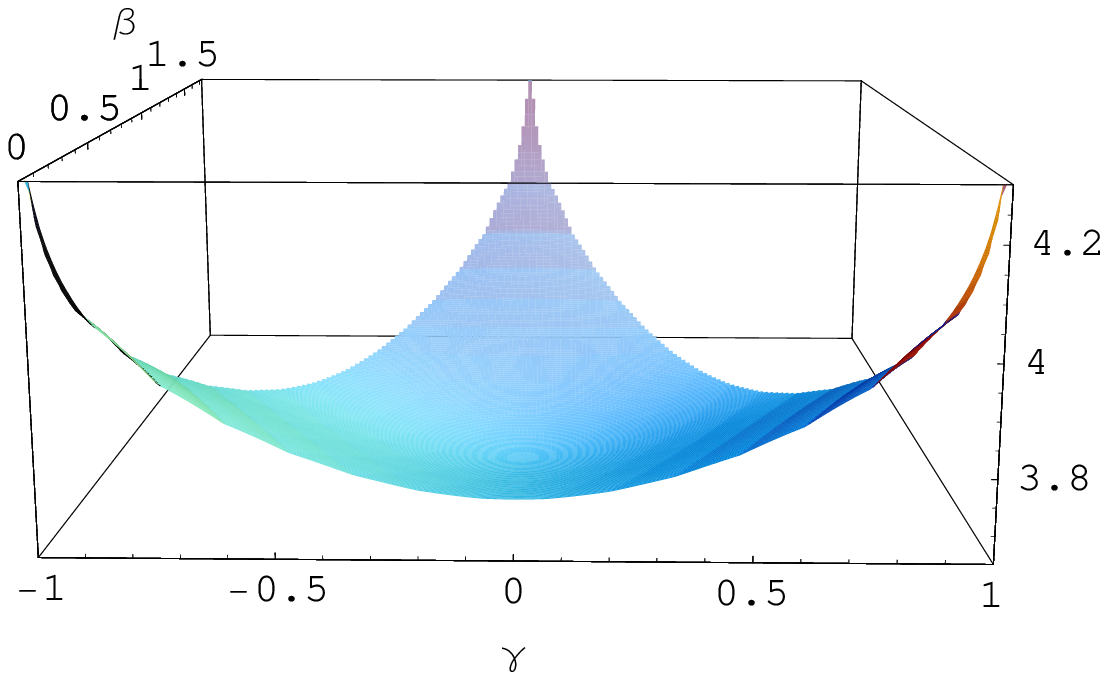}}
    %\hspace{0.5cm}
    \subfigure[\label{trianglefig}]{\includegraphics[width=6cm]{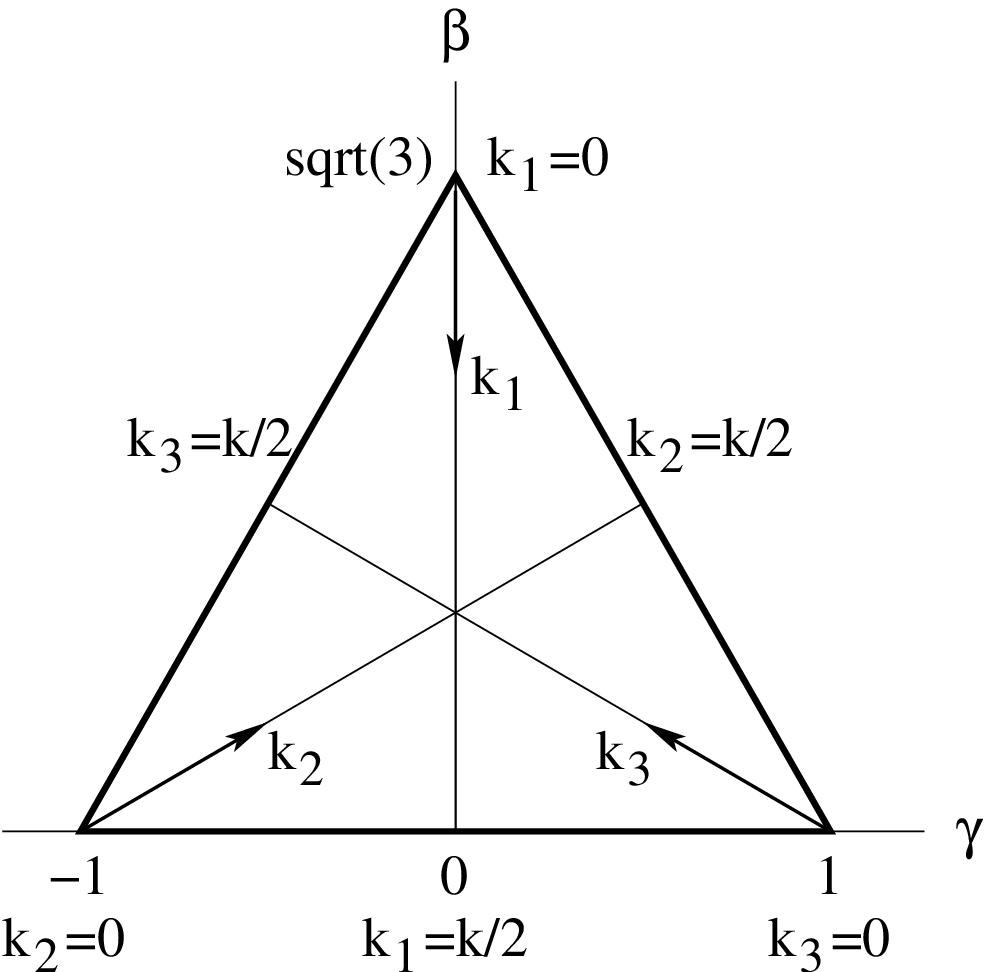}}
    \caption{(a) The bispectrum (\ref{3pcorr}), (\ref{three point}) without
    the overall $(2\pi)^3 \gd^3(\sum_s \vc{k}_s)$ factor and divided by the
    square of the power spectrum (\ref{powerspec2f}) as defined in
    (\ref{deffNL}), plotted as a function of
    the relative size of the three momenta. The sum of the momenta is chosen
    as $(k_1+k_2+k_3) = 3 k_*$ with $k_*$ fixed by choosing $\gD t_* = 50$. 
    The values for the parameters are $\tge = -\tget^\parallel = 0.05$, 
    $\tget^\perp = 0.2$, $\gc = 0.01$, $(\sqrt{2\tge}/\gk)V_{111}/(3 H^2) 
    = -\tgx^\perp = -(\sqrt{2\tge}/\gk)V_{221}/(3 H^2) 
    = (\sqrt{2\tge}/\gk)V_{222}/(3 H^2) = 0.003$ (as well as $c=3$, although 
    the dependence on $c$ is negligible). (b) An explanation
    of the triangular domain used, defined in (\ref{plotvars}), with
    $k\equiv k_1+k_2+k_3$.}
    \end{figure}

After having discussed the various momentum limits, we finally
show the full dependence on the relative magnitude of the momenta
of the bispectrum divided by the square of the power spectrum in
figure~\ref{3pcorrfig}, where we did not use any analytic
approximations for the integrals. To be precise, it is actually
the bispectrum given in (\ref{3pcorr}) and (\ref{three point}),
without the overall $(2\pi)^3 \gd^3(\sum_s \vc{k}_s)$ factor (but
taking into account the relation between the momenta that the
$\gd$-function implies), divided by the sum of products of power
spectra (\ref{powerspec2f}) with different momenta, as follows:
    \beq\label{deffNL}
    \tf_\mathrm{NL} \equiv
    \frac{\langle \tgz^1 \tgz^1 \tgz^1 \rangle(k_1,k_2,k_3)}
    {\left[\langle \tgz^1 \tgz^1 \rangle(k_1) \langle \tgz^1 \tgz^1 \rangle(k_2)
    +\langle \tgz^1 \tgz^1 \rangle(k_1) \langle \tgz^1 \tgz^1 \rangle(k_3)
    +\langle \tgz^1 \tgz^1 \rangle(k_2) \langle \tgz^1 \tgz^1 \rangle(k_3)
    \right]/3}.
    \eeq
This quantity can be seen as a momentum-dependent version of the
$f_\mathrm{NL}$ parameter often used in the literature (see e.g.\
\cite{ngreview}).\footnote{There is a difference of a factor of
order unity between $\tf_\mathrm{NL}$ and $f_\mathrm{NL}$ even in
the equal momentum limit, caused partly by the difference between
$\gz$ and the gravitational potential $\gF$ which was used in the
original definition.} We now choose $k_*$ to be the mode that
crossed the horizon 50 e-folds before the end of inflation (i.e.\
we set $\gD t_* = 50$). The function $\tf_\mathrm{NL}$ depends on
the three scalars $k_1, k_2, k_3$, but we can plot it in a
two-dimensional triangular domain if we fix their sum, which we do
by setting $(k_1+k_2+k_3)/k_* = 3$. This convenient way of
plotting the three-point correlator in a triangle, clearly
demonstrating its symmetries, was introduced in \cite{sf}, and is
illustrated in figure~\ref{trianglefig}.\footnote{Note, however,
that in \cite{sf} a different normalisation factor was used.} The
quantities on the axes are
    \beq\label{plotvars}
    \gg \equiv 2 \, \frac{k_2 - k_3}{k_1 + k_2 + k_3},
    \qquad\qquad
    \gb \equiv - \sqrt{3} \, \frac{k_1 - k_2 - k_3}{k_1 + k_2 + k_3}.
    \eeq
At the vertices of the triangle one of the three momenta is equal
to zero. Lines of constant $k_s$ are parallel to the sides of the
triangle (a different side for each $s=1,2,3$) and $k_s$ increases
linearly perpendicular to these. At the side itself the
corresponding momentum is equal to half the total sum,
$(k_1+k_2+k_3)/2$. In the centre of the triangle all momenta have
equal length.

The plot in figure~\ref{3pcorrfig} has been made for all
first-order slow-roll parameters equal to $0.05$, except
$\tget^\perp=0.2$ and $\gc = 0.01$, and all second-order slow-roll
parameters equal to $0.003$. We see that there is a dependence on
the relative magnitude of the momenta. Though not visible in the
figure, this dependence is strongest very near the vertices of the
triangle, which is the limit of (\ref{3pcvertex}), where for this
specific example the value $9.4$ is reached. Of course
logarithmically the region near the vertices covers an infinite
range of magnitudes in momentum ratios. (The fact that the result
is largest in the squeezed momentum limit agrees with the findings
of \cite{BCZ}.) The value at the centre is $3.7$. Assuming that a
naive extrapolation of this result at the end of inflation to the
time of recombination is allowed, so that the quantity plotted is
indeed comparable to the observable $f_\mathrm{NL}$, we see that
this model does produce sufficient non-Gaussianity to be
detectable with the Planck satellite. To compare this plot with
the one for the single-field case in \cite{sf} one should keep in
mind that there an additional factor of $(2\tge+\tget^\parallel)$
was left out (and there are some differences in the momentum
normalisation factor, but that does not change the magnitude
much), so that the multiple-field result is indeed about two
orders of magnitude larger.

\subsection{Comparison with semi-analytic calculation using quadratic 
potential} 
\label{realmodel}

Of course it may be argued that the approximate model considered
here is not very realistic, with all slow-roll parameters constant
with time (in particular $\tget^\perp$ and $\gc$). We should also
stress that the calculation here was made under the assumption
that $\gc>0$, which is not true for all models. While we will
study more realistic models in great detail in a future
publication \cite{RSvTnum}, both semi-analytically and purely
numerically without any approximations, for direct comparison here
we present a bispectrum calculation using the Green's function
formalism outlined in section~\ref{generalsec}. We have
investigated a simple two-field model with a quadratic potential
$V = \frac{1}{2} m_1^2 \gf_1^2 + \frac{1}{2} m_2^2 \gf_2^2$ with
$m_1 = 1\cdot 10^{-5} \gk \inv$ (the overall mass magnitude can be
freely adjusted to fix the amplitude of the power spectrum). The
analytic solution (\ref{linsol}) is used as the linear source term
in (\ref{bia2}), the super-horizon Green's function is then
calculated from (\ref{Green}) and (\ref{2fA}), and the bispectrum computed 
from these using (\ref{generalfkkreal}) and (\ref{bA}). We find that for a 
mass ratio $m_2/m_1 = 9$ and initial conditions $\gf_1 = \gf_2 = 13 \gk\inv$
we get relatively large non-Gaussianity: with all momenta equal,
that is, at the centre of the triangle, the ratio of the
bispectrum to the square of the power spectrum is, in the slow-roll limit,
    \beq
    \tf_\mathrm{NL} = 1.8,
    \eeq
where we have taken horizon crossing to be 58 e-folds before the
end of inflation. The ratio of the contribution from the
$I$-integrals to that of the $J$- integrals is $74$. This confirms
our assertion that the integrated secular terms (the $\bA$ term in
(\ref{generalfkkreal})) subsequent to horizon crossing dominate
the contributions to the bispectrum. We note also that the
spectral index in this model is $0.93$, which is observationally
acceptable. While the investigation of the quadratic model is
preliminary at this stage, it is clear that large non-Gaussianity
($f_\mathrm{NL}$ greater than unity) can be obtained in a real
multiple-field inflation model.

Even though the slow-roll parameters are definitely not constant
in the quadratic model, we find that the expressions in the
previous subsection can be used to make a useful approximation of
the numerical result. In this case, for example, to estimate
$\tget^\perp$ we use a representative value and adjust $\gD t$ to
reflect the region of support where its value is significant. To
compare to the quadratic potential result, here we have taken
$(\tget^\perp)^2 = 0.73$ (its maximum value) and $\gD t = 1.0$
(full width at half maximum), and use the limit $2(\tget^\perp)^2
\gD t$ given below (\ref{centreratio}). From this we estimate the
value $1.5$, which is relatively close to the numerical value. This
seems to indicate that the constant slow-roll, analytic results of
the previous subsection\footnote{Except expression (\ref{spec})
for the spectral index, which is a poor approximation in this
case.}, while in principle unrealistic, can be used to get a first
estimate of the amount of non-Gaussianity even in real models with
varying slow-roll parameters.\footnote{Note added: After our
investigation of this example in an earlier version of this paper,
the non-Gaussianity produced by a quadratic potential was also
considered by the authors of \cite{lyth alabidi}. They conclude
that $f_{NL}\ll 1$ using the `$\delta N$ formalism' and
extrapolating results from a potential with (almost) equal masses.
They have not computed the general case of unequal masses and
assume deviations from the (almost) equal mass case to be small.
In the case of almost equal masses, which is effectively
single-field, we agree that non-Gaussianity is small. However,
already in \cite{vantent} it was quantitatively shown that even at
linear order additional leading-order effects arise in the case of
unequal masses from the effective coupling between the fields
caused by the bending of the trajectory in field space. Moreover,
in the model we consider here, the dominant non-Gaussianity is
caused by a relatively large $\tget^\perp$, so that a naive
slow-roll order counting is not valid, a situation where the
calculation of \cite{lyth alabidi} as we understand it is not applicable.}

\subsection{Discussion}
\label{Discuss}

Before we conclude, a couple of points regarding the consistency
of our approach need to be discussed. The first is an inherent
limitation of our method in capturing all possible sources of
non-Gaussianity, since, by using the linear perturbation solutions
to source our non-linear equations, we are implicitly neglecting
all non-linear interactions up to horizon crossing. We believe
that this accounts for the small discrepancy between the momentum
dependence of our single-field three-point correlator \cite{sf}
and that obtained from the tree-level action calculations of
\cite{maldacena} when $k_1\approx k_2\approx k_3$, whereas in the
limit $k_1,k_2 \mg k_3$ the two correlators agree exactly. We
surmise that the super-horizon non-linear effects described by our
method and the horizon-crossing effects we are missing are of
comparable magnitude for single-field inflation in the equal
momentum limit.

For multiple-field inflation, however, the situation is very
different. We can see this by using the quantitative results above
to interpret our key integral expression for the three-point
correlator (\ref{generalfkkreal}). In the case where
multiple-field effects are large (as indicated by the behaviour of
$\tget^\perp$) it is the perturbation of the long-wavelength
evolution term in the integrand of (\ref{generalfkkreal})
(represented by $\bA^{(0)}_{abc}$ and absent in the single-field
case) that dominates over perturbations of the stochastic source
term which contains the linear perturbations (represented by
$\bX^{(1)}_{amc}$; the term that would, in principle, be
influenced by these horizon-crossing effects). For example, in the
case considered in figure~\ref{3pcorrfig} the contribution of the
$\bX^{(1)}_{amc}$ term, which is given by the $J$-integrals in
(\ref{three point}), is 62 times smaller than the contribution
from the $\bA^{(0)}_{abc}$ term at the vertices of the triangle,
and 177 times smaller at the centre. But would the
$\bX^{(1)}_{amc}$ term be similarly enhanced when taking into
account non-linear effects at horizon crossing in the
multiple-field case? This seems unlikely, given that horizon
crossing is only a short transition, while the large effects of
the other term are caused by a buildup over a significant time
interval. Moreover, this question appears to have been answered
definitively in the negative by recent work \cite{seery}.
Generalising \cite{maldacena} for multiple-field inflation, though
only up to horizon crossing, it shows that these extra
contributions remain of the order of small slow-roll parameters,
just as in the single-field case. In that sense, the papers
\cite{maldacena,seery} are important null results which clarify
that our approach focusing on non-linear super-horizon effects
will indeed capture the main non-Gaussian contributions from
multiple-field inflation models.

The second point regards the possible influence of loop
corrections to the stochastic picture for generating and evolving
inflationary perturbations. It is generally accepted within the
cosmological community that quantum fluctuations can be considered
classical for modes which have crossed the horizon, and we
explicitly make such an assumption here by using classical random
fields to set up initial conditions for long wavelengths via the
source terms in (\ref{sources}). The long-wavelength evolution is
then followed by using the classical equations of motion. A
natural question to be asked is whether loop corrections might
play a role in the super-horizon evolution. Recently, the question
was addressed in \cite{weinberg} for a single inflaton field plus
a number of non-interacting massless scalar fields. A theorem was
proved about the growth of loop effects and it was shown that for
the theories mentioned loop effects were determined at horizon
crossing and were subdominant. Since the conditions of the theorem
imply $\tget^\perp=0$, these results are not directly applicable
to the kind of models considered in this paper. However, even if
loop effects were to grow with time in such models, they would
still need to dominate over the classical growth that these models
can exhibit in order to interfere with the classical picture for
the evolution of the perturbations we have developed here.
Nevertheless, a definitive answer to such matters requires further
investigation.

\section{Conclusions}
\label{conclsec}

In this paper we have investigated non-Gaussianity in
multiple-field inflation using the formalism of \cite{gp2,
formalism}, emphasizing analytic calculations. That formalism is
based on fully non-linear equations for long wavelengths, with
stochastic source terms taking into account the short-wavelength
quantum fluctuations. For analytic calculations an expansion of
the relevant equations in perturbation orders is necessary.
However, it is much easier to derive the perturbed equations at
second order directly from the non-linear equation of motion for
$\gz_i$ than from perturbing the original Einstein equations. Of
course, in a fully numerical investigation no expansion in
perturbation orders has to be made; this will be explored in
future work.

We derived two main results in this paper. The first is the
general solution for the bispectrum, (\ref{generalbispec}) with
(\ref{generalfkk}) or (\ref{generalfkkreal}). Even though this is
an integral expression, it will be relatively simple to evaluate
in a semi-analytic calculation and it yields the full momentum
dependence.  To achieve this one only needs solutions for the
homogeneous background quantities in the inflation model, for the
linear perturbation variable $Q^\mathrm{lin}$ around horizon
crossing, and for the homogeneous Green's function, as well as
expressions for the spatial derivatives of the various coefficient
functions. The latter can all be computed analytically from the
constraint equations (\ref{constr1})--(\ref{constr3}); for the
general two-field case all relevant expressions were given
explicitly in \cite{formalism} and this paper. Computing the
bispectrum is then just a question of performing a few time
integrals. An accurate semi-analytic treatment will be the subject
of a forthcoming paper \cite{RSvTnum}, though we do provide the
results of a slow-roll calculation for a quadratic potential here. In
the present paper, however, we have emphasized an example where we
could proceed purely analytically.

In the second part of the paper we studied two-field slow-roll
inflation, with the strong leading-order approximation that all
slow-roll parameters are constant. In this case we could work out
the bispectrum explicitly analytically (apart from a few integrals
that had to be done numerically, although we found analytic
approximations in certain limits), which is the other main result
of this paper, equation (\ref{three point}). We found that in this
two-field case the bispectrum can easily be two orders of
magnitude larger than in the single-field case, due to the
continued buildup of non-Gaussianity on super-horizon scales
caused by the influence of the isocurvature mode on the adiabatic
perturbation. We note that even though the presence of isocurvature
perturbations during inflation is crucial, it is not mandatory
that they survive afterwards. In fact they feed into the adiabatic
perturbation and can disappear by the end of inflation (as in the cases studied
here). On the other hand, if any isocurvature modes do persist at the end of
inflation, their fate will depend on the details of reheating and further
evolution.

The bispectrum divided by the square of the power spectrum, which
can be seen as a momentum-dependent generalisation of the
$f_\mathrm{NL}$ observable, can be $f_\mathrm{NL}\approx
\cO(1)$--$\cO(10)$, or even larger in extreme cases. If a
straightforward extrapolation of this result at the end of
inflation to the time of recombination is justified, a subject
which still needs to be studied in more detail, this means that
the Planck satellite, and to a lesser extent even the WMAP
satellite, should be able to confirm or rule out certain classes
of multiple-field inflation models. Finally we want to stress that
beyond estimating the amplitude of the bispectrum, we also give
its explicit momentum dependence. While this dependence is rather
flat for momenta of comparable size, there is a significant
difference between more extreme momentum limits.

We believe this paper is a significant step towards providing
quantitative and testable predictions of non-Gaussianity from
multiple-field inflation. Nevertheless there are still a number of
issues that remain to be investigated in more detail, and on which
we are working for future publications. In the first place, we
will apply our general solution for the bispectrum to more
realistic inflation models, particularly those strongly motivated
by fundamental theory. This will require a semi-analytic
treatment, but because we are dealing with integral equations, we
believe that the strong slow-roll approximations presented here
will actually provide reasonable analytic estimates of the exact results.
As a first step we presented here the results of the semi-analytic slow-roll
treatment of an explicit two-field model with a quadratic
potential. This will be investigated in more detail in
\cite{mf2,RSvTnum}, but the results confirm the fact that
non-Gaussianity can be large in multiple-field inflation models,
and that our analytic approximations provide a good estimate.
Next, it is of course important to study the further evolution of
non-Gaussianities after inflation through recombination to the
present day (see \cite{ngreview, BMR} for some work in this
direction). In this paper we restricted ourselves to computing
only the bispectrum of the adiabatic component of $\gz$, even
though we have the solution for all components. In future work we
will investigate isocurvature and mixed bispectra as well.
Finally, we will test our results with a purely numerical
implementation of our formalism, which can also be applied to
non-slow-roll models where our analytic approximations fail.  In
this case, the real-space realisations for $\gz$ that result allow
for other measures of non-Gaussianity to be determined, not just
the three-point (or higher) correlator. After the disappointing
results for single-field inflation, primordial non-Gaussianity is
now back as an important quantitative tool for confirming or
ruling out multiple-field inflation models, offering an exciting
new window on the early universe as observations continue to
improve over the next 5--10 years.

\section*{Acknowledgements}

We thank the organisers of the ``The Origin of the Primordial
Density Perturbation'' workshop in Lancaster, UK, in March 2005
for organising such a stimulating meeting, especially regarding
non-Gaussianity, and where we presented the first version of the
results of section~\ref{expl2fsec}. This research is supported by
PPARC grant PP/C501676/1.

\end{document}